\documentclass[preprint,12pt,authoryear]{elsarticle}

\usepackage[margin=1in]{geometry}
\usepackage[T1]{fontenc}
\usepackage{mathptmx}
\usepackage{courier}
\usepackage{amssymb,amsmath}
\usepackage{booktabs,multirow,graphicx}
\usepackage{subcaption}
\usepackage{setspace}
\usepackage[mathlines]{lineno}
\usepackage{microtype}
\usepackage{xurl}
\usepackage[hidelinks]{hyperref}
\usepackage[nameinlink,capitalise,noabbrev]{cleveref}

\journal{Journal of Informetrics}
\newcommand{\tabnote}[1]{\par\smallskip\begin{minipage}{\linewidth}\footnotesize #1\end{minipage}}

\begin{document}

\begin{frontmatter}

\title{What Changed in Engineering Writing After Generative AI Chat Assistants:
Evidence from 1.8 Million Articles}

\author{Taner Cokyasar\corref{cor1}}
\ead{tcokyasar@tamu.edu}
\cortext[cor1]{Corresponding author.}

\affiliation{organization={Department of Engineering Technology and Industrial
  Distribution, Texas A\&M University},
  city={College Station}, postcode={77843}, state={TX}, country={USA}}

\begin{abstract}
This study asks what changed in engineering articles after
conversational language models became publicly available on November~30, 2022, and
develops methods for measuring such change in a field without a uniform abstract
index. It assembled 1,839,962 articles published from 2015 to 2025 in 256 engineering
journals, tracked twelve measures of wording, style, and article structure, and
compared each with the projection of its own pre-2023 trend. Building the monthly
series showed that records dated only by year fall on January~1 and create false
January peaks unless removed. In abstracts, 31 words associated with language model
output appeared 82\% more often than projected (55.4 against 27.0 per 10,000 words in
2025), while four control words of the same formal register fell 16\%. Words that
soften a claim, such as \textit{may}, fell 13\%, words that strengthen one, such as
\textit{clearly}, rose 20\%, and long words rose 11\%. Author counts, reference lists,
page counts, and publication volume stayed within 4\% of trend. The change appears in
all fourteen subject categories, and false release dates, dropping any one journal or
word, and word lists chosen from the data support it. A random sample of 12,866 full
texts from the same corpus, including subscription articles and weighted back to
it, shows the same words rising 99\% in article bodies. Statements disclosing AI
writing assistance, measured apart from mentions of language models, went from none
to 3.7\% in 2025, four in five of them at one publisher. The design is descriptive,
not causal.
\end{abstract}

\begin{keyword}
scholarly writing \sep generative artificial intelligence \sep large language models
\sep bibliometrics \sep interrupted time series \sep engineering
\end{keyword}

\end{frontmatter}

\section{Introduction}
\label{sec:introduction}

On November~30, 2022, a conversational language model became freely available to
anyone with a web browser \citep{openai2022chatgpt}. Within months such models were
listed as coauthors \citep{stokelwalker2023}, editors were reporting machine-written
submissions \citep{conroy2023}, publishers had issued disclosure policies
\citep{cope2023ai}, and a substantial share of surveyed researchers said they used
the tools \citep{vannoorden2023}. Much of what is claimed about the effect on
academic writing, however, rests on anecdote, on small hand-coded samples, or on
detection tools whose error rates are poorly known and that misjudge authors writing
in an additional language \citep{liang2023bias}. This paper asks a narrower question:
what measurably changed in the engineering research literature after that date?

Engineering is a useful field for this analysis, for three reasons. It is large, so
monthly series are precise. It spans subject areas with different writing conventions, so a change can be
checked across fields. It has also received less attention than two other fields:
biomedicine, where PubMed collects the abstracts of the literature in one index, and
computer science, where many papers appear in full on the arXiv preprint server.
Engineering has no such source of abstracts or full texts, so this study had to assemble
both, and that shapes its design.

Beyond its findings for engineering, the paper makes three contributions to how such
change is measured, each of which applies to any field. First, it
combines two sources. Monthly series are built from the \emph{corpus}, meaning every
research article that the study's journals published from 2015 to 2025, as described
by open bibliographic records. A sample of full texts is then weighted back to the
corpus, so that results from abstracts can be checked against the body of the same
articles in a field without a uniform abstract index.
Second, it shows that records carrying only a publication year, which bibliographic
sources assign to January~1, produce false January peaks in monthly series, and it
removes them. Third, it separates statements that an AI tool was used to write an
article from mentions of language models as a subject or a citation; counting the
latter overstates disclosure.

The study measures twelve things, in three groups, and the results follow the same
order. Each was chosen because the tools could plausibly change it, and each is
introduced here with the change it was meant to detect.

Four come from article abstracts. The first is how often words associated with
language model output appear (marker words, such as \textit{delve},
\textit{intricate}, \textit{showcase}, and \textit{underscore}): if authors draft or
edit with these tools, the words the models over-use should become more common. The
marker words come mainly from the words that \citet{kobak2025} and
\citet{liang2024reviews} found language models over-use; \Cref{sec:methods} lists all
31 and their sources. The second is how often authors soften a claim with qualifying
words, such as \textit{may} and \textit{might}, and how often they strengthen one with
emphatic words, such as \textit{clearly} and \textit{substantially}; text drafted or
polished by a model may state its claims more firmly, with fewer qualifiers and more
emphasis. The third is how long and how varied the words are, which would rise if the
tools replace plain words with longer and less common ones. The fourth is
readability, which could move either way: the tools are often used to make prose
clearer, but longer words make it harder to read. Readability also stands in for
grammatical quality, which cannot be measured on published articles because
copyediting removes most errors before publication.

Five come from the bibliographic record. If the tools make writing faster, more
articles could be published, so publication volume could rise; if they let one
person do work that once needed several, the number of authors and institutions per
article could fall; and if they make the literature easier to search and summarize,
reference lists could grow. Title form (length, and whether a colon splits the title
in two) is included because language models favor two-part titles
\citep{sebo2026titles,igras2026colons}. The share of
articles that are open access is not expected to respond to writing tools. It is
tracked because a change in how articles are published alters which publishers make
up the literature, and such a shift could be mistaken for a change in writing.

Three require the full text and are measured on a sample. Marker words in the body
of the article test whether a change seen in abstracts, short texts written to a
fixed brief, extends to the article itself. Explicit statements that an AI tool was
used to write the article, which publishing-ethics guidance began to call for in
2023 \citep{cope2023ai}, show how much of that use authors acknowledge. How the article is
assembled (figures, tables, equations, and length) would change if the tools let
authors produce and present more analysis in the same time, giving longer articles
with more figures and tables.

Each measure is compared with the projection of its own pre-2023 trend rather than
with its past average, because scientific writing drifts steadily and a
before-and-after difference would mistake that drift for change.

The study's design is descriptive. For each measure, the study fits the path the measure
followed from January 2015 to November 2022 and extends that path beyond the release
date; the departure from trend is the gap between what was then observed and this
projection. A departure shows that something changed, but it cannot be attributed to
a cause, because there is no control group: no comparable body of writing that the
tools left untouched. The paper also makes no claim about any individual article.
Every measure is a frequency over many articles and cannot be used to label one.

\section{Background}
\label{sec:background}

The first scholarly response to public chat assistants concerned authorship.
Editorials held that a tool cannot take responsibility for a work and so cannot be an
author \citep{stokelwalker2023,thorp2023}, position statements settled on disclosure
plus human accountability \citep{cope2023ai}, and a literature on acceptable use
followed \citep{hosseini2024,haman2024,dergaa2023}. In the indexed literature, the
tools came to be acknowledged or cited rather than listed as authors
\citep{gorraiz2025invisible}. This work set the terms of the
debate but measured little; it rests on editorial experience, surveys, and
demonstrations that machine-written abstracts can pass human inspection
\citep{else2023abstracts}.

For the first measure, marker words, two approaches exist. Classifying individual
texts as machine written is unreliable at scale, because detectors misclassify
competent non-native academic prose at high rates \citep{liang2023bias}. Measuring
aggregate change avoids that problem. \citet{kobak2025} compared the frequency of
every word in millions of PubMed abstracts with a projection of that word's own
earlier trend and found that specific words became abruptly more common after 2022.
Two studies by Liang and colleagues estimated, from the whole vocabulary, the share of
model-modified text in scientific papers \citep{liang2024papers} and in conference
peer reviews \citep{liang2024reviews}. In both approaches the words are an output of the analysis, not an
input. \citet{kousha2025llms} took the other route, as this study does: they tracked a
fixed set of such words across several bibliographic databases and in full texts, and
found their use still rising through 2025. \citet{hubarava2026drift} applied an
interrupted time series to preprint abstracts and titles in several fields and found
longer words, lower readability, and accelerating use of colons in abstracts after
2022. \citet{bao2025shifts} reported the same pattern in more than 800,000 arXiv
abstracts: words that language models prefer became more common, lexical complexity
rose, and readability fell, most of all in computer science.
\citet{alsudais2025readability} found readability falling across arXiv categories
after the release, and \citet{alani2026stylistic}, comparing a fixed list of marker
words with a control vocabulary in biomedical abstracts, found that marker use more
than doubled while the control vocabulary declined slightly. An earlier study of
fabricated papers showed that machine-mediated text leaves lexical traces that predate
2022 \citep{cabanac2021tortured}.

Disclosure has been measured directly only recently. \citet{he2026policies} classified
the AI policies of 5,114 journals and estimated AI-assisted writing in 5.2 million
abstracts published from 2021 to mid-2025. About 70\% of the journals had a policy,
most of them requiring disclosure, yet estimated use rose about as fast in journals
with a policy as in those without one, and of about 75,000 full texts published from
2023, 76 (0.1\%) disclosed the use of an AI writing tool.

This study differs from that work in five ways. The first is the field. The studies
above draw on PubMed, on preprint servers whose texts are uniformly available, or on
whole databases divided into broad areas, and none of them reports engineering
journals separately; in \citet{he2026policies}, for example, engineering is part of
the physical sciences. The second is the comparison. Several of these studies compare
the years before and after the release \citep{kousha2025llms,bao2025shifts,alsudais2025readability},
which can mistake steady drift for change. This study, like \citet{kobak2025} and
\citet{hubarava2026drift}, compares each measure with the projection of its own
trend, and it adds control words of the same register, false release dates, and an
adjustment for the pandemic months. The third is breadth. Most earlier studies measure
wording or style alone; this study places wording beside the bibliographic record of
the same articles (authors, institutions, references, pages, publication volume,
titles, and open access), which shows whether a change in wording came with a change
in how research is produced and published. The fourth is the full-text sample.
Earlier full-text analyses use open-access collections \citep{kousha2025llms} or the
copies a script could download \citep{he2026policies}, which over-represent
open-access publishers. This study draws articles at random from all the articles of
each journal and year, including subscription articles, and weights them back to the
corpus. The fifth is the disclosure measure, which separates a statement that a tool
was used on the text from a mention of language models as a topic, and which was
checked by reading every stored document it flags. The study also documents a data
problem that monthly designs share: records that carry only a publication year fall
on January~1 and create false peaks.

The other measures have been studied as long-run trends rather than in relation to a
specific date. Qualifying and emphatic language are established categories in the
study of academic writing \citep{hyland1998}, and readability formulas have a long
history \citep{flesch1948}. Author counts and reference lists have grown for decades,
and the literature itself grows several percent per year \citep{bornmann2015growth},
which is why a simple before-and-after comparison cannot be read directly. The
segmented (interrupted time series) regression used here
\citep{wagner2002segmented,bernal2017its} applies the same projection logic to every
measure, with standard errors that allow for correlation between neighboring months
\citep{newey1987}.

\section{Data}
\label{sec:data}

As defined in the introduction, the corpus is the set of all research articles published from
January 2015 to December 2025 by a fixed list of engineering journals, together with
each article's bibliographic record and, where available, its abstract. The window
ends in 2025 because records for 2026 were still incomplete when the data were
collected. The year left out is likely to matter. In a Wiley survey of 2,430
researchers in August 2025, 84\% reported using AI tools in their work, up from 57\% a
year earlier \citep{wiley2025explanaitions}, and in a survey of more than 10,000
researchers in March 2026, nearly half reported frequent use of AI for gathering
information and editing \citep{springernature2026state}. The tools also improved
quickly: in one year, frontier models gained about 30 percentage points on Humanity's
Last Exam, a benchmark of difficult expert questions, and one reached gold-medal level
at the 2025 International Mathematical Olympiad \citep{aiindex2026}. The results here
therefore describe the first three years after the release, not the state of writing
in 2026.

The journals come from Google Scholar Metrics \citep{scholarmetrics}, which lists the
twenty journals with the highest h5-index in each subject category. Using a
published ranking means any reader can rebuild the list. Fourteen engineering
categories were used: general engineering and computer science; food science and
technology; manufacturing and machinery; materials engineering; mechanical
engineering; sustainable energy; computer networks and wireless communication; game
theory and decision science; transportation; automation and control theory; power
engineering; structural engineering; civil engineering; and operations research.
Each listed journal was matched to its OpenAlex record by title. Conference proceedings were removed, and a journal
listed in two categories was kept in the one where it ranks higher, leaving 256
journals. The time series use the 229 of these journals that published at least
twenty articles in every year of the window, so that a journal's founding does not
look like a change in the literature.

All English-language works these journals published in the window that OpenAlex
\citep{priem2022openalex} classified as articles were collected in September 2026,
giving 1,839,962 articles; OpenAlex's article type excludes editorials, letters, and
errata but includes some reviews. OpenAlex has become a common source for such
studies, although its metadata on languages, document types, and affiliations remain
incomplete \citep{forchino2026openalex}. Abstracts come
from OpenAlex; where OpenAlex has none, the abstract indexed by Semantic Scholar was
used when available. About 8.5\% of the articles in the 229 journals carry January~1
as their publication date because only the year is recorded; they are kept in annual
figures but left out of monthly series, where they would otherwise pile into January.

The four abstract measures need abstract text, which publishers deposit unevenly
(\Cref{tab:coverage}). Three publishers, Elsevier, Springer, and the Royal Society of
Chemistry, deposit few abstracts openly and together hold 46\% of the articles.
Overall, 68.0\% of the more than 1.8 million articles have an abstract of at least forty words,
the minimum used here, and 14.4\% of those abstracts come from Semantic Scholar.

\begin{table}[htbp]
\caption{Share of Articles with an Abstract of at Least Forty Words, by Publisher.}
\label{tab:coverage}
\centering
\singlespacing\small
\begin{tabular}{lrr}
\toprule
Publisher & Articles & With abstract \\
\midrule
Elsevier & 749,523 & 35.9\% \\
Multidisciplinary Digital Publishing Institute & 333,294 & 99.3\% \\
Institute of Electrical and Electronics Engineers & 265,095 & 97.6\% \\
American Chemical Society & 131,155 & 99.7\% \\
Wiley & 122,360 & 98.9\% \\
Springer & 64,065 & 21.6\% \\
Royal Society of Chemistry & 32,228 & 3.2\% \\
Taylor \& Francis & 29,232 & 98.9\% \\
All others & 113,010 & 85.8\% \\
\midrule
All 256 journals & 1,839,962 & 68.0\% \\
\bottomrule
\end{tabular}
\end{table}

\section{Methods}
\label{sec:methods}

Every measure is treated the same way. Its monthly values from 2015 to 2022 define a
trend, the trend is projected forward, and the result is the gap between what was
observed from 2023 onward and that projection.

The model has the following terms. Time is counted in months: $t$ is the number of
months since January 2015, so $t=0$ is January 2015, $t=12$ is January 2016, and so
on, and the value of a measure in month $t$ is $y_t$. The last month before
the release, November 2022, is denoted $T_0$. The indicator $D_t=1$ for
months after $T_0$ and zero, otherwise. The indicator $C_t=1$ during the
acute phase of the COVID-19 pandemic, March 2020 to June 2021, and zero otherwise.
Research and publishing ran under unusual conditions in those months: editorial
decisions came faster \citep{sun2023expedited} and patterns of collaboration shifted
\citep{ni2025disruptions}. Some series moved temporarily, for example references per article (\Cref{fig:meta-refs}). The
indicator lets the level of each measure shift during those sixteen months without
bending the trend fitted to the rest of 2015 to 2022; it is zero after June 2021 and
so plays no part in the projection. Separately, $m$ denotes the month of the year,
from $m=1$ for January to $m=12$ for December, and is used only for seasonal patterns:
for $m$ from 2 to 12, the indicator $S_{m,t}=1$ when month $t$ falls in month $m$ of
its year (for example, $S_{2,t}=1$ in every February), and zero otherwise. These
indicators absorb regular seasonal patterns in publishing, with January as the
reference month. The
coefficients are $\beta_0$ to $\beta_4$ and $\gamma_2$ to $\gamma_{12}$, and
$\varepsilon_t$ is the error. The model is
\begin{equation}
  y_t = \beta_0 + \beta_1 t + \beta_2 D_t + \beta_3 (t - T_0) D_t
        + \beta_4 C_t + \sum_{m=2}^{12} \gamma_m S_{m,t} + \varepsilon_t .
  \label{eq:its}
\end{equation}
Here $\beta_1$ is the monthly trend before the release, $\beta_2$ an immediate change
in level, and $\beta_3$ a change in that trend afterward. Articles published from
December 2022 to February 2023 were written before the release and are left out of
the estimation. The projection is \Cref{eq:its} with $\beta_2$ and $\beta_3$ set to
zero, and the reported departure is the mean observed value from March 2023 to
December 2025 divided by the mean projected value over the same months, minus one.
Standard errors allow for correlation across twelve neighboring months
\citep{newey1987}.

Frequencies are pooled within each month: the number of occurrences of the words in
question in all abstracts published that month, divided by the total number of words
in those abstracts, times 10,000.

The marker list holds 31 words. Some words were listed with inflected forms, and each word is counted together with
its listed forms, so that \textit{underscore}, \textit{underscores}, and
\textit{underscoring} are one word in every table and figure. The other forms counted
are \textit{aligning}, \textit{fostering}, \textit{harnessing}, \textit{intricacies},
\textit{leveraging}, \textit{meticulously}, \textit{seamlessly}, \textit{showcasing},
and \textit{unveiling}. The words are
\textit{additionally}, \textit{align}, \textit{commendable}, \textit{comprehensive},
\textit{consequently}, \textit{crucial}, \textit{cutting-edge}, \textit{delve},
\textit{elevate}, \textit{foster}, \textit{furthermore}, \textit{harness},
\textit{innovative}, \textit{intricate}, \textit{landscape}, \textit{leverage},
\textit{meticulous}, \textit{multifaceted}, \textit{notably}, \textit{noteworthy},
\textit{nuanced}, \textit{paradigm}, \textit{pivotal}, \textit{realm},
\textit{robust}, \textit{seamless}, \textit{showcase}, \textit{significant},
\textit{tapestry}, \textit{underscore}, and \textit{unveil}. Twenty-four come from
published lists of words that language models over-use: the excess words that
\citet{kobak2025} found in biomedical abstracts and the adjectives and adverbs that
\citet{liang2024reviews} report for peer reviews. Seven, \textit{cutting-edge},
\textit{furthermore}, \textit{landscape}, \textit{paradigm}, \textit{robust},
\textit{significant}, and \textit{tapestry}, appear in neither source and were added by
the author. Neither source used a fixed list as its method; both found these words by
searching the whole vocabulary. Using their findings as a fixed list in another field
is therefore this study's choice, and the result is also checked with lists chosen
from the data, described with the results.

Four further words, \textit{moreover}, \textit{vital}, \textit{holistic}, and
\textit{state-of-the-art}, serve as controls. They belong to the same formal register
as the marker words but do not appear in published lists of model-favored words. They
are counted separately and are not part of the marker measure. If marker words rose
only because academic prose in general became more formal, the controls would rise
too.

For the full vocabulary, every word is grouped with its inflected forms using a
standard English stemming algorithm, so that, for example, \textit{enhance},
\textit{enhanced}, and \textit{enhancing} count as one word. For each word $w$ and year
$k$, let $f_{w,k}$ be its frequency per million words in a sample that takes at most
400 abstracts from each journal in each year, so that no journal's subject matter
dominates. A straight line fitted to $f_{w,k}$ over 2015 to 2022 gives projected
values $\hat{f}_{w,k}$ for later years. The excess ratio of word $w$ is the observed
mean over 2023 to 2025 divided by the projected mean over the same years,
\begin{equation}
  R_w = \frac{\sum_{k=2023}^{2025} f_{w,k}}{\sum_{k=2023}^{2025} \hat{f}_{w,k}} .
  \label{eq:excess}
\end{equation}
A value of one means the word followed its own trend; two means it appeared twice as
often as the trend implied. Words occurring fewer than 400 times in the sample are
left out, and ratios are reported only when the projection is at least one per million
words, because a near-zero denominator makes any ratio meaningless.

The remaining abstract measures use every word of the abstract, not the marker list.
Qualifying words are words that soften a claim, such as \textit{may}, \textit{might},
and \textit{suggests}; emphatic words strengthen one, such as \textit{clearly},
\textit{substantially}, and \textit{significantly}. Both follow \citet{hyland1998} and
are counted per 10,000 words. Word length is the share of words with nine or more
letters. Word variety is the number of distinct words in each block of 100
consecutive words, averaged over the blocks of an abstract; counting within
fixed-size blocks keeps longer abstracts from looking less varied merely because they
are longer. Readability is the Flesch-Kincaid grade level \citep{flesch1948}, which
rises with syllables per word and with words per sentence. The abstracts supplied by
OpenAlex keep word order but lose punctuation, so sentence length cannot be measured;
it is set to 22 words for every abstract. The grade therefore reflects syllables per
word only, and only its change over time is interpreted.

The bibliographic measures come directly from each article's record: the numbers of
authors, institutions, and references, the page count, the number of articles
published per month, whether the article is open access, and two features of the
title, its length in words and whether it contains a colon. A colon splits a title into
a main title and a subtitle; such titles are a long-studied feature of academic
writing \citep{lewison2005title,hartley2007colons}, and language models produce them
more often than human authors do \citep{sebo2026titles,igras2026colons}.

\section{Results}
\label{sec:results}

\subsection{Marker words}
\label{sec:results-vocab}

Marker words departed from their trend more than any other measure. From 2015 to
2022 their frequency rose slowly, from 22.3 to 25.9 per 10,000 words, and the annual
values stayed within 2.5\% of the fitted trend (\Cref{tab:yearly}). From 2023 the two
separated: by 2025 the observed frequency was 55.4 against a projection of 27.0.
Averaged over March 2023 to December 2025, marker words appeared 81.8\% more often
than projected. Over the same months the four control words appeared 15.6\% less
often than their projection.

\begin{table}[htbp]
\caption{Marker-Word Frequency per 10,000 Words, Observed and Projected.}
\label{tab:yearly}
\centering
\singlespacing\small
\begin{tabular}{lrrr}
\toprule
Year & Observed & Projected & Ratio \\
\midrule
2015 & 22.3 & 21.9 & 1.02 \\
2016 & 22.6 & 22.4 & 1.01 \\
2017 & 23.1 & 22.9 & 1.01 \\
2018 & 23.0 & 23.4 & 0.98 \\
2019 & 23.4 & 23.9 & 0.98 \\
2020 & 23.9 & 24.3 & 0.98 \\
2021 & 25.2 & 24.9 & 1.01 \\
2022 & 25.9 & 25.5 & 1.02 \\
\midrule
2023 & 35.6 & 25.8 & 1.38 \\
2024 & 51.2 & 26.5 & 1.94 \\
2025 & 55.4 & 27.0 & 2.05 \\
\bottomrule
\end{tabular}
\tabnote{Note: Ratio is observed divided by projected; 2.05 means the words appeared
2.05 times as often as the pre-2023 trend implied. Values are averages of monthly
values. Before 2023 the projected column is the fitted trend.}
\end{table}

\begin{figure}[htbp]
\centering
\includegraphics[width=0.9\linewidth]{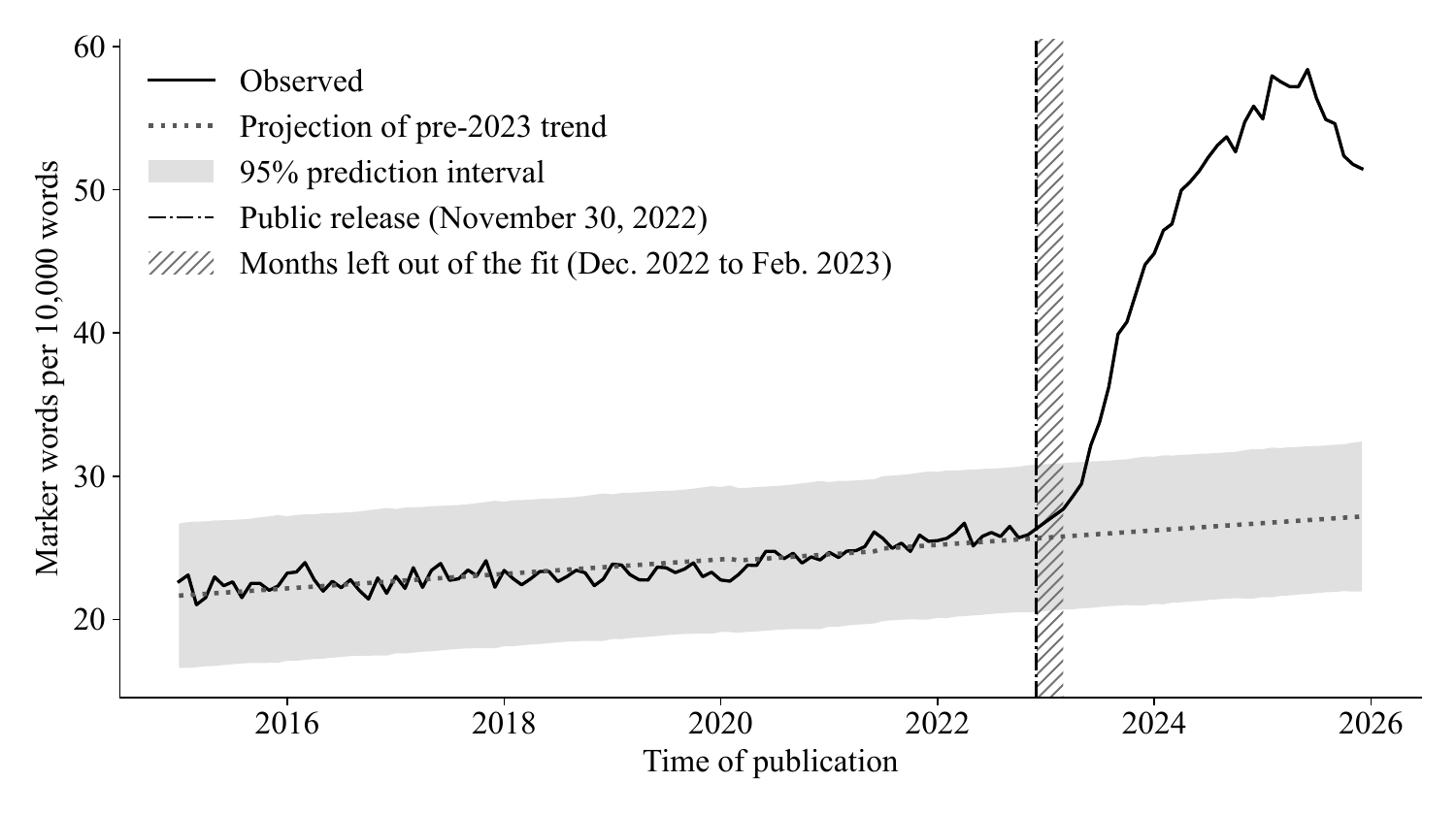}
\caption{Monthly Marker-Word Frequency in Abstracts and the Projection of Its
Pre-2023 Trend.}
\label{fig:marker}
\end{figure}

\Cref{fig:marker} shows the monthly series. The rise was gradual rather than
immediate: the series left the projection band in mid-2023 and climbed through 2024,
as expected when writing, review, and publication take months. Accordingly, the model
finds a clear change in trend ($p < 0.001$) but no statistically significant jump at
the release date ($p = 0.14$). The figure therefore shows the observed series and the
projection only. The month-to-month pattern is regular: across calendar months the
pre-2023 averages range only from 23.2 to 24.1 per 10,000 words once records dated
only by year are excluded; including them would raise every January by 5\% over the
other months.

The series peaked in June 2025 at 58.4 and declined to 51.5 by December. The decline
is not in the data's composition: mean abstract length (189.4 and 189.5 words) and the
share of abstracts taken from Semantic Scholar (5.0\% and 4.5\%) were flat between the
first and second halves of 2025. It is in the words. From the first to the second half
of 2025, \textit{additionally} fell 28\%, \textit{crucial} 21\%, \textit{innovative}
21\%, and \textit{showcase} 34\%, while \textit{robust} rose 32\% and
\textit{paradigm} 78\%. The mix of favored words is shifting, which is consistent
with changes in the tools or in how authors use them, though this study cannot
separate the two. A similar pattern appeared in arXiv abstracts, where words publicly
identified as typical of language models, such as \textit{delve}, fell soon after they
were publicized while others kept rising, which \citet{geng2025coevolution} attribute
to authors adapting how they use the tools. Emphatic words kept rising and qualifying words kept falling over
the same months.

A rise in a chosen list is weak evidence on its own; the question is whether it is
confined to the expected words. \Cref{tab:markers} shows that it largely is.
\textit{Underscore} appeared 24.0 times as often as projected and nine other marker
words between 2.9 and 5.8 times, while the controls stayed between 0.81 and 1.14.
The boundary is graded: common connectives in the list such as
\textit{furthermore} (1.16) and \textit{consequently} (1.15) rose only modestly. The
word most associated with machine-written prose in popular accounts, \textit{delve},
is rare in engineering abstracts: it rose from 0.7 to 8.1 occurrences per million
words, with a projection below one per million, so it has no ratio.

\begin{table}[htbp]
\caption{Marker and Control Words Against Their Projected Trends, per Million Words.}
\label{tab:markers}
\centering
\singlespacing\small
\begin{tabular}{lrrrr}
\toprule
Word & 2015--2022 & Projected & 2023--2025 & Ratio \\
\midrule
\multicolumn{5}{l}{\emph{Ten largest departures among marker words}} \\
\textit{underscore} & 3.8 & 4.5 & 107.5 & 24.04 \\
\textit{intricate} & 7.4 & 8.7 & 50.6 & 5.83 \\
\textit{meticulous} & 1.8 & 3.5 & 14.3 & 4.05 \\
\textit{pivotal} & 11.6 & 15.8 & 60.3 & 3.82 \\
\textit{align} & 11.7 & 14.0 & 52.5 & 3.74 \\
\textit{realm} & 3.2 & 3.8 & 13.7 & 3.60 \\
\textit{nuanced} & 2.4 & 3.7 & 12.5 & 3.36 \\
\textit{foster} & 11.3 & 13.3 & 39.0 & 2.94 \\
\textit{showcase} & 6.9 & 14.2 & 41.5 & 2.93 \\
\textit{multifaceted} & 3.4 & 4.1 & 11.9 & 2.90 \\
\midrule
\multicolumn{5}{l}{\emph{Control words}} \\
\textit{holistic} & 24.0 & 32.3 & 36.8 & 1.14 \\
\textit{vital} & 53.0 & 75.0 & 75.5 & 1.01 \\
\textit{state-of-the-art} & 129.1 & 200.4 & 195.3 & 0.97 \\
\textit{moreover} & 496.6 & 620.6 & 499.8 & 0.81 \\
\bottomrule
\end{tabular}
\tabnote{Note: Each word includes its listed inflected forms. Frequencies come from a
sample of at most 400 abstracts per journal per year.}
\end{table}

\begin{figure}[htbp]
\centering
\includegraphics[width=0.95\linewidth]{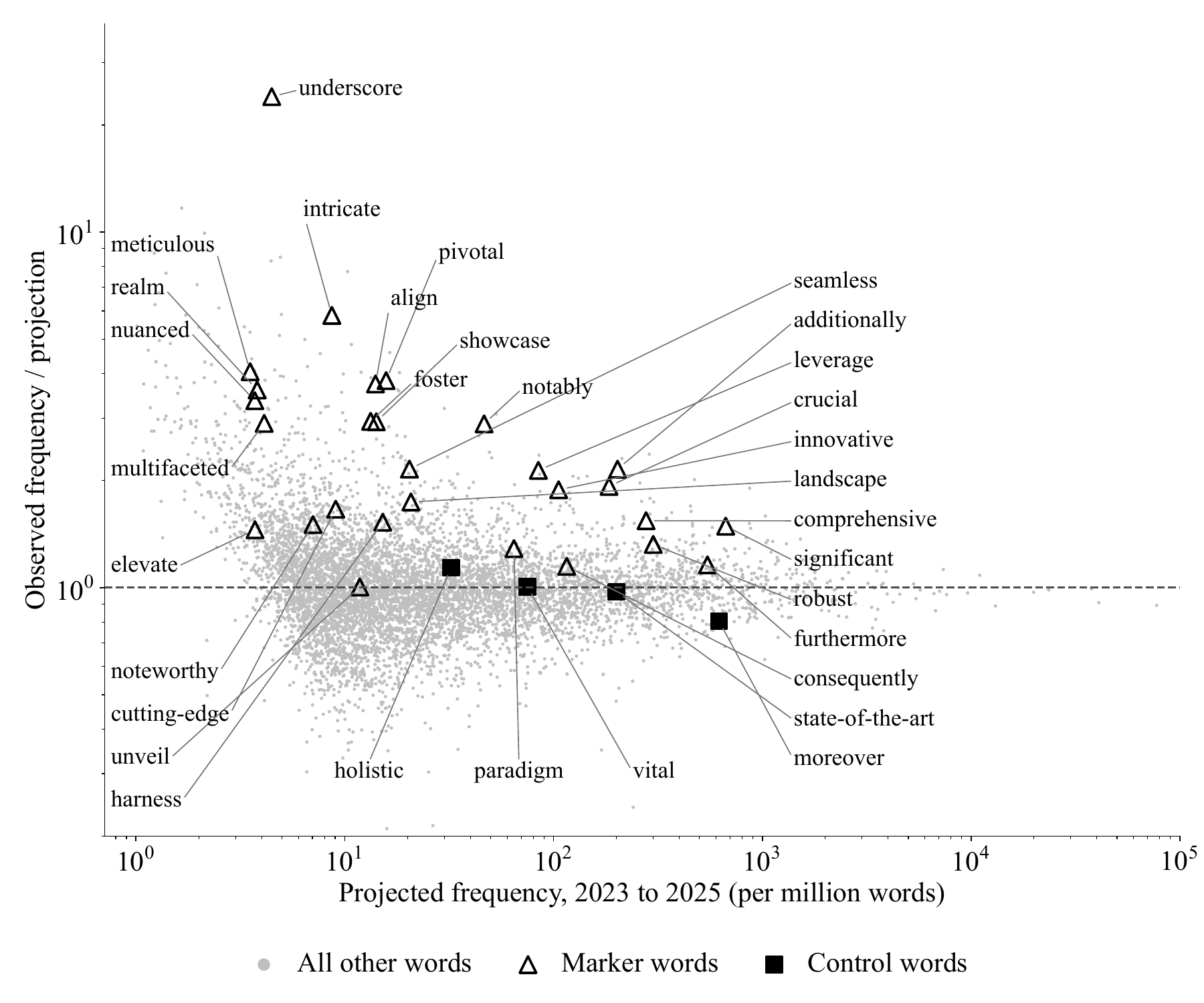}
\caption{Excess Ratio of Every Word Occurring at Least 400 Times, with Marker and
Control Words Labeled.}
\label{fig:excess}
\end{figure}

\Cref{fig:excess} places the marker and control words within the whole vocabulary.
Most marker words lie above the bulk of other words, and the controls lie at one,
with ordinary words. The marker words are not a separate cluster, however: many other
words also rose, so the list captures part of a broader shift rather than all of it.

The change reached the typical article. Pooling all abstracts from all journals and
years on each side of the release (\Cref{tab:distribution}), the share of abstracts
with no marker word at all fell from 67.0\% to 45.6\%, the median rose from zero to
42.0 per 10,000 words, and every upper percentile rose by more than half.

\begin{table}[htbp]
\caption{Marker Words per 10,000 Words in Individual Abstracts, Before and After the
Release.}
\label{tab:distribution}
\centering
\singlespacing\small
\begin{tabular}{lrr}
\toprule
Statistic & Before & After \\
\midrule
Share of abstracts with no marker word & 67.0\% & 45.6\% \\
Median & 0.0 & 42.0 \\
75th percentile & 49.3 & 76.6 \\
90th percentile & 75.8 & 134.5 \\
99th percentile & 169.5 & 260.9 \\
Mean & 24.1 & 50.1 \\
\bottomrule
\end{tabular}
\tabnote{Note: Each abstract contributes one value; all journals and all years on each
side are pooled. Before: published to November 2022; after: from March 2023.}
\end{table}

\subsection{Robustness of the marker-word result}

\Cref{tab:robustness} reports five kinds of check. First, the model was refitted
with a false release date in November 2018, 2019, or 2020, using only data from before
the real release; the departures are 4\% to 7\%, which reflects slight curvature in
the earlier trend and is a small fraction of the 82\% found at the real date. Second,
removing each of the 228 journals with abstract text in turn (one of the 229 has
none) leaves the estimate between 79.4\% and 83.1\%, and removing each of the 31 marker
words in turn leaves it between 77.3\% and 97.2\%. Third, the sample was varied: giving each journal a fixed weight in every
month, so that changes in which journals publish most cannot drive the result,
gives 72.4\%; using only journals whose abstracts are nearly complete, only abstracts
from OpenAlex, all 256 journals, or the records dated only by year keeps the estimate
between 80.4\% and 83.9\%. Fourth, dropping the level-change term from \Cref{eq:its}
gives 68.9\%.
Fifth, the word list was varied. The seven words added by the author are common in
engineering prose for reasons unrelated to language models (\textit{significant} alone
accounts for 27.7\% of marker occurrences before the release, \textit{furthermore} for
20.6\%, and \textit{robust} for 11.4\%), and together they make up 62.6\% of those
occurrences. They dilute the measure rather than drive it: the 24 words taken from the
published sources depart by 130.1\% from their trend, and the seven added words by
47.3\%. In addition, two word lists were chosen from the data by a fixed rule: words in use in all
fourteen categories before 2023 that rose above their own projection in at least
twelve of them, excluding function words. The held-out list was chosen using only
abstracts published before April 2024 and evaluated only on later months; the
disjoint list shares no word with the marker or control lists. Both depart far more
than the marker list, as expected for words selected because they rose, but the
held-out list shows that words chosen before the evaluation period kept rising
afterward.

\begin{table}[htbp]
\caption{Robustness of the Marker-Word Departure.}
\label{tab:robustness}
\centering
\singlespacing\small
\begin{tabular}{llr}
\toprule
Check & Variant & Departure \\
\midrule
Main estimate & 31 marker words, 229 journals & $+81.8\%$ \\
\midrule
\multirow{3}{*}{False release date}
  & November 2018 & $+3.6\%$ \\
  & November 2019 & $+5.1\%$ \\
  & November 2020 & $+6.7\%$ \\
\midrule
Leave one journal out & Range across 228 journals & 79.4 to 83.1\% \\
\midrule
\multirow{6}{*}{Sample}
  & Fixed journal weights & $+72.4\%$ \\
  & Journals with nearly complete abstracts & $+83.9\%$ \\
  & OpenAlex abstracts only & $+80.4\%$ \\
  & Without journals whose abstracts were restored & $+82.5\%$ \\
  & All 256 journals & $+82.3\%$ \\
  & Including records dated January~1 & $+83.2\%$ \\
\midrule
Model & No level-change term & $+68.9\%$ \\
\midrule
\multirow{6}{*}{Word list}
  & Control words & $-15.6\%$ \\
  & 24 words from published sources & $+130.1\%$ \\
  & 7 words added by the author & $+47.3\%$ \\
  & Leave one marker word out & 77.3 to 97.2\% \\
  & Held-out list (40 words) & $+156.0\%$ \\
  & Disjoint list (40 words) & $+121.5\%$ \\
\bottomrule
\end{tabular}
\tabnote{Note: False-release fits use only data before the real release. For 81
journals the OpenAlex abstracts were collected in a second pass, after the rest of
their records; the row ``without journals whose abstracts were restored'' leaves
those journals out.}
\end{table}

\begin{figure}[htbp]
\centering
\includegraphics[width=0.8\linewidth]{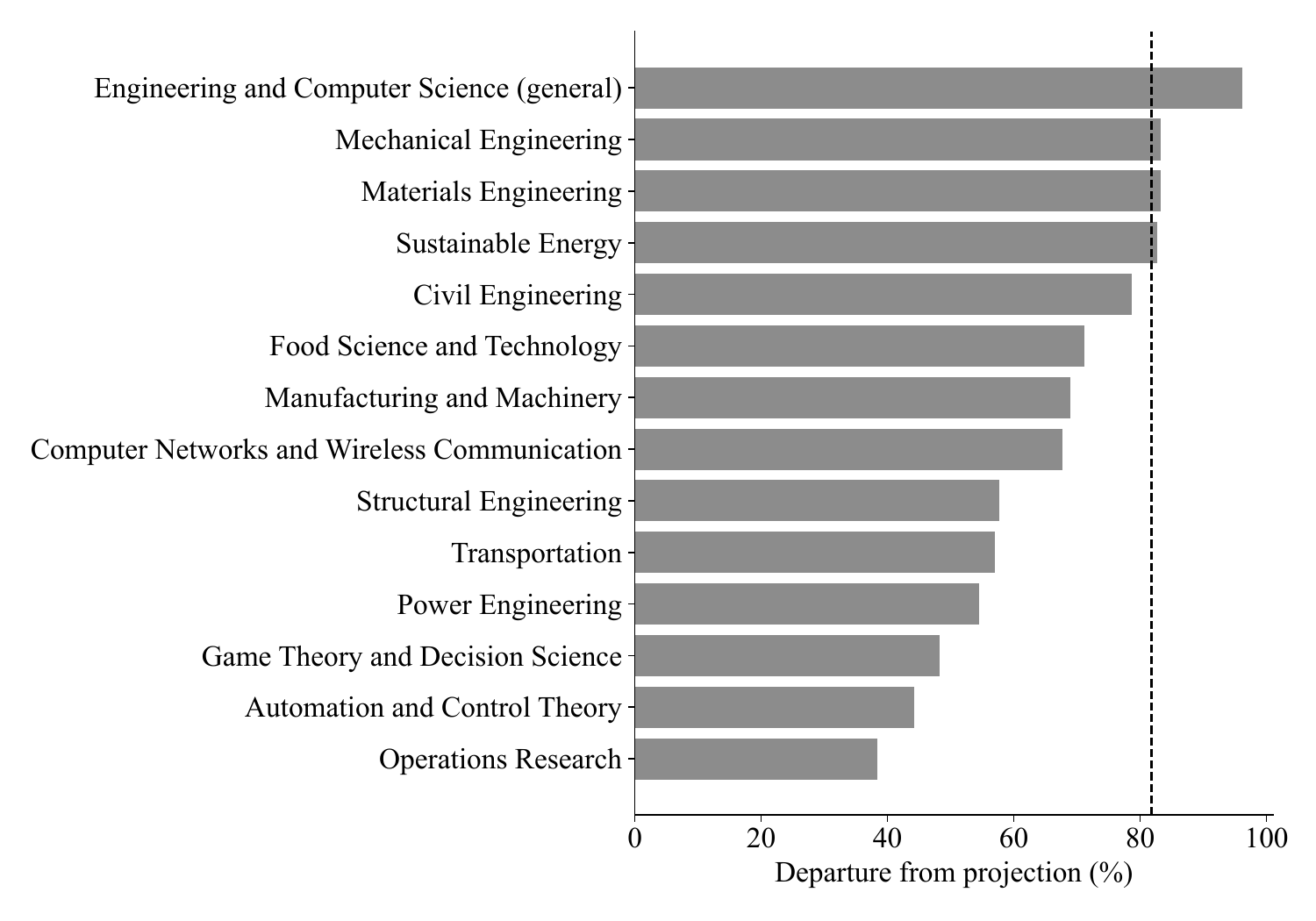}
\caption{Marker-Word Departure by Subject Category.}
\label{fig:category}
\end{figure}

A change in what these journals publish would be expected to concentrate in some
fields. Instead, all fourteen categories depart upward and significantly
(\Cref{fig:category}), from 38.4\% in operations research to 96.2\% in general
engineering and computer science. Fields that share little in topic, authors, or
publisher moved in the same direction at the same time.

Splitting the series by the country of the corresponding author shows both groups
rising: 57.5\% for authors based in countries where English is the main language of
research and 87.7\% for others. Country is a poor stand-in for an author's language
background, so this split is descriptive only.

\subsection{Qualifying and emphatic words, word length, and readability}

\begin{figure}[htbp]
\centering
\begin{subfigure}[t]{0.49\linewidth}
\centering
\includegraphics[width=\linewidth]{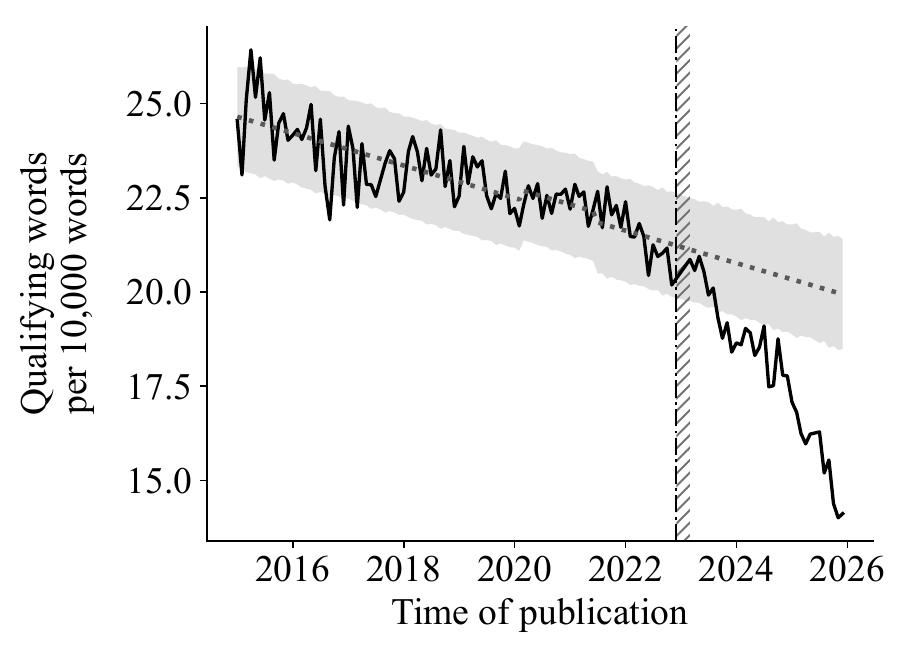}
\caption{Qualifying words}
\label{fig:style-hedge}
\end{subfigure}\hfill
\begin{subfigure}[t]{0.49\linewidth}
\centering
\includegraphics[width=\linewidth]{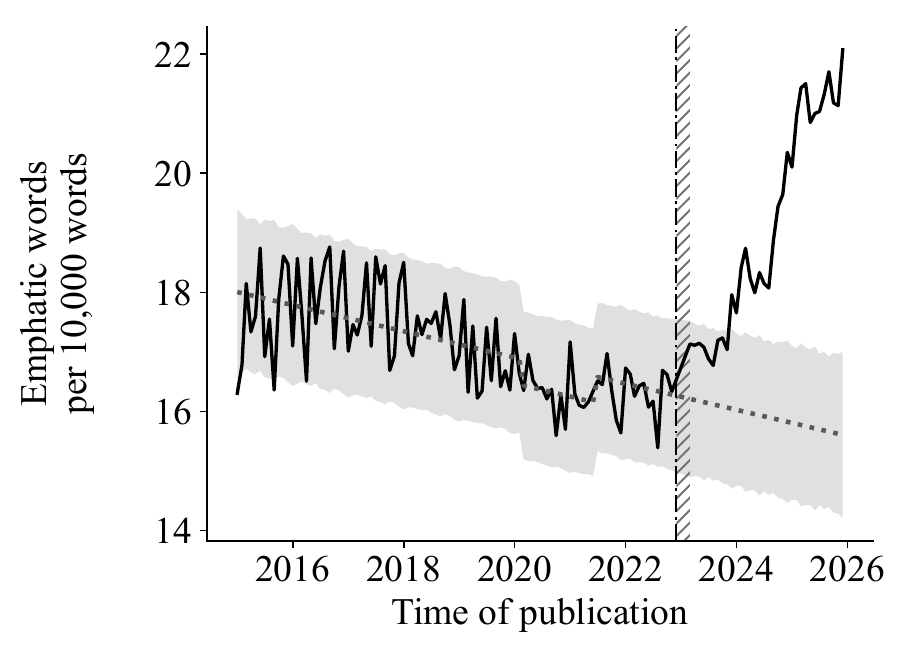}
\caption{Emphatic words}
\label{fig:style-booster}
\end{subfigure}
\par\medskip
\begin{subfigure}[t]{0.49\linewidth}
\centering
\includegraphics[width=\linewidth]{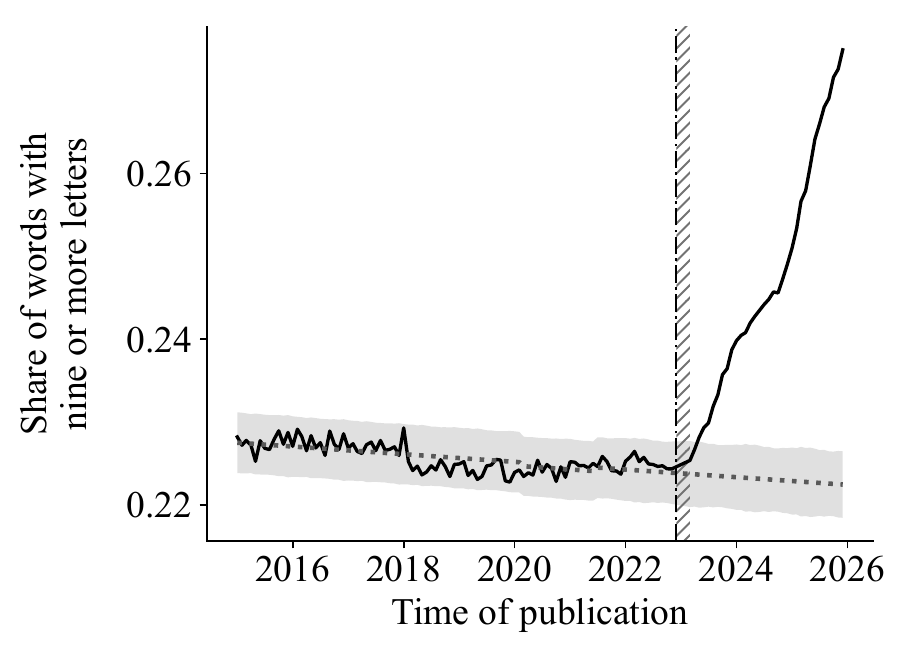}
\caption{Long words}
\label{fig:style-long}
\end{subfigure}\hfill
\begin{subfigure}[t]{0.49\linewidth}
\centering
\includegraphics[width=\linewidth]{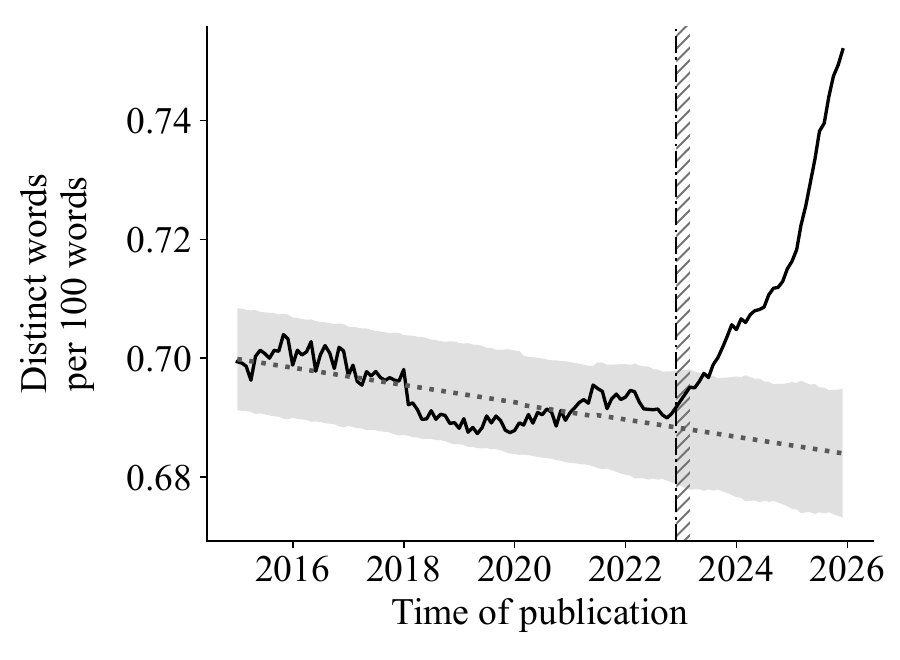}
\caption{Word variety}
\label{fig:style-ttr}
\end{subfigure}
\par\medskip
\begin{subfigure}[t]{0.49\linewidth}
\centering
\includegraphics[width=\linewidth]{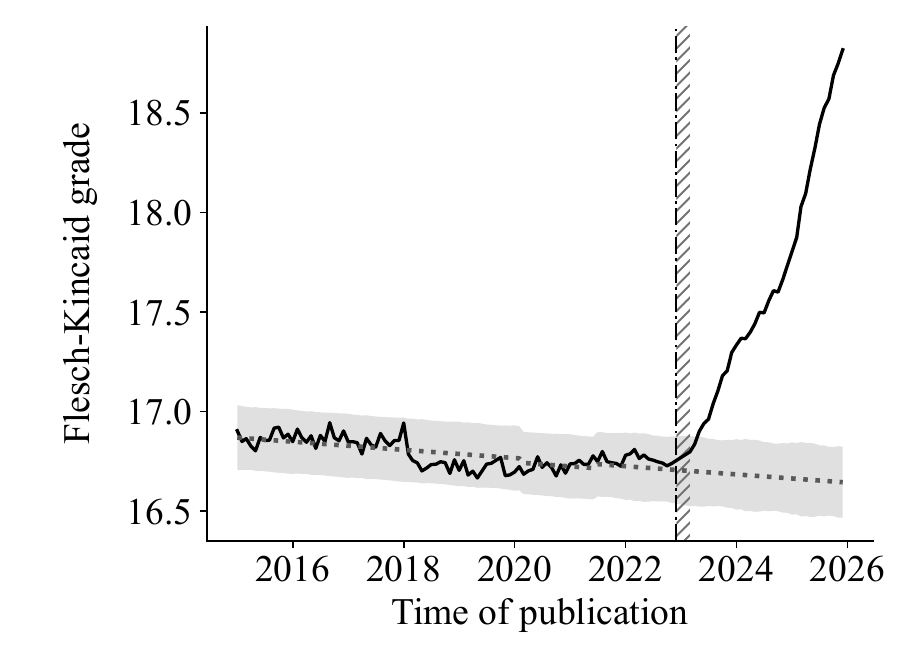}
\caption{Readability grade}
\label{fig:style-fk}
\end{subfigure}\hfill
\begin{subfigure}[t]{0.49\linewidth}
\centering
\includegraphics[width=\linewidth]{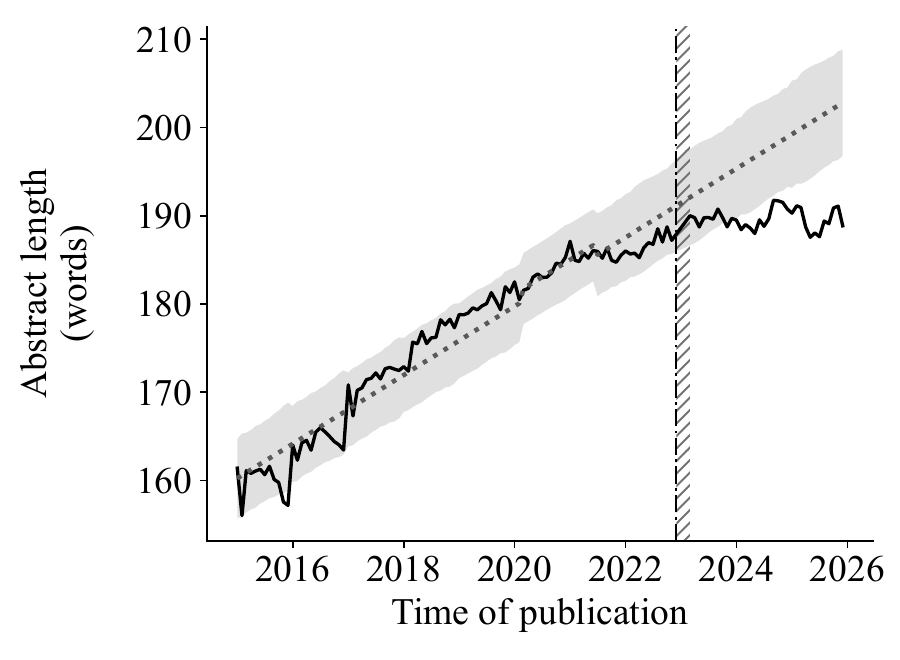}
\caption{Abstract length}
\label{fig:style-len}
\end{subfigure}
\par\medskip
\includegraphics[width=0.95\linewidth]{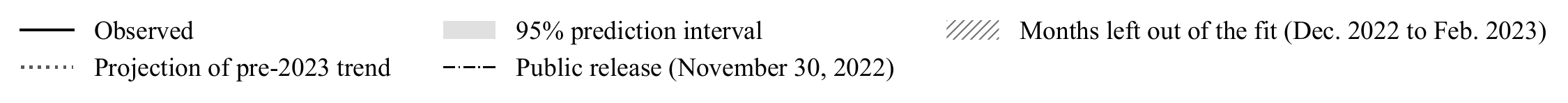}
\caption{Abstract Measures and the Projections of Their Pre-2023 Trends. The step in each projection from March 2020 to June 2021 is the
pandemic adjustment in \Cref{eq:its}.}
\label{fig:style}
\end{figure}

The other abstract measures moved together with marker words (\Cref{tab:its},
\Cref{fig:style}). Qualifying words, the usual way scientific writing signals
uncertainty, fell 13.0\% below projection (\Cref{fig:style-hedge}), while emphatic
words rose 20.1\% (\Cref{fig:style-booster}). Words became longer: the share of words
with nine or more letters rose 10.8\% (\Cref{fig:style-long}), word variety 4.3\%
(\Cref{fig:style-ttr}), and the readability grade 5.9\% (\Cref{fig:style-fk}).
Abstracts became slightly shorter than their rising trend implied ($-4.0\%$;
\Cref{fig:style-len}).

\subsection{Bibliographic measures}

\begin{figure}[htbp]
\centering
\begin{subfigure}[t]{0.49\linewidth}
\centering
\includegraphics[width=\linewidth]{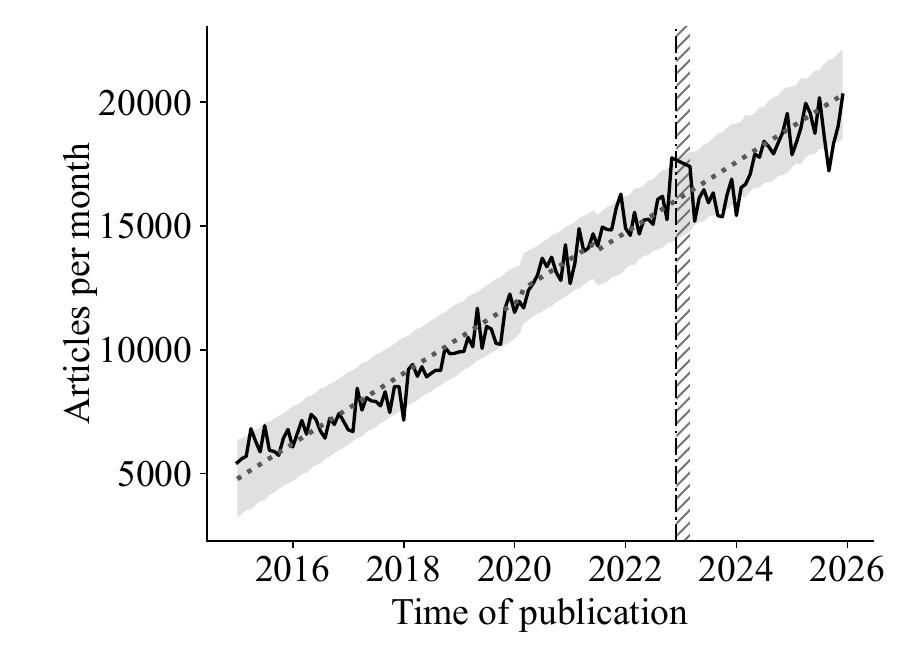}
\caption{Articles per month}
\label{fig:meta-articles}
\end{subfigure}\hfill
\begin{subfigure}[t]{0.49\linewidth}
\centering
\includegraphics[width=\linewidth]{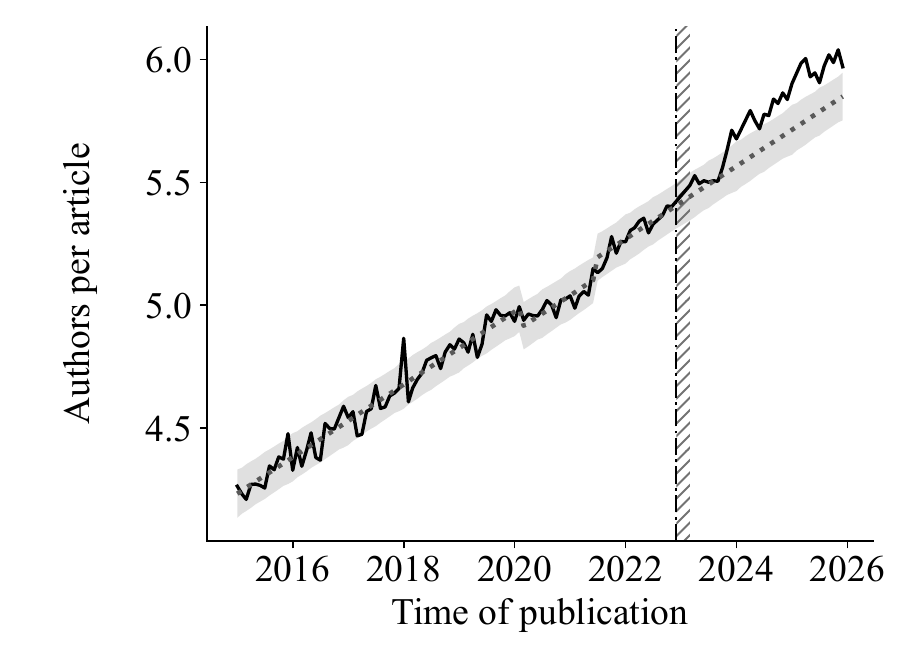}
\caption{Authors per article}
\label{fig:meta-authors}
\end{subfigure}
\par\medskip
\begin{subfigure}[t]{0.49\linewidth}
\centering
\includegraphics[width=\linewidth]{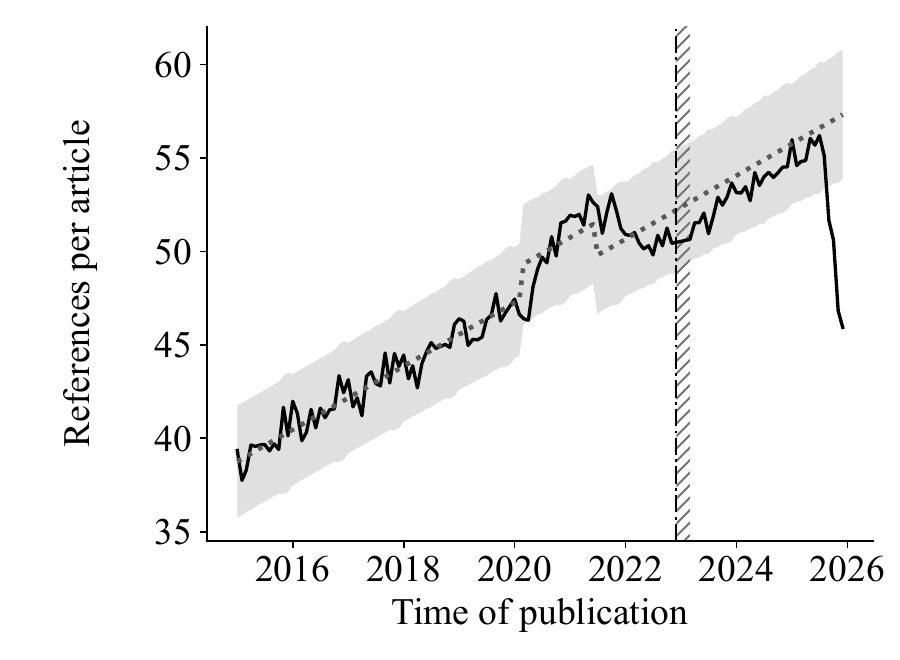}
\caption{References per article}
\label{fig:meta-refs}
\end{subfigure}\hfill
\begin{subfigure}[t]{0.49\linewidth}
\centering
\includegraphics[width=\linewidth]{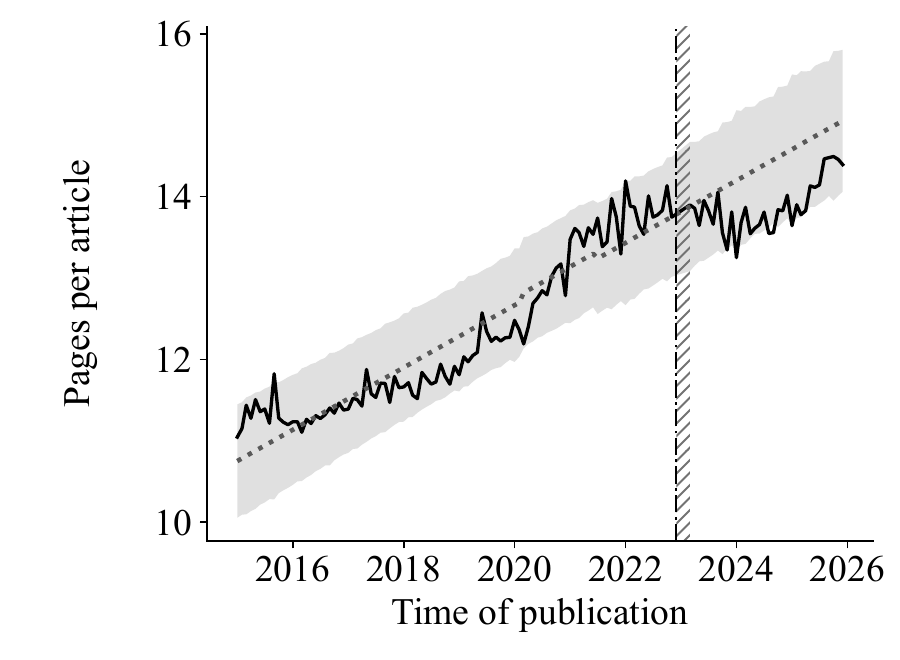}
\caption{Pages per article}
\label{fig:meta-pages}
\end{subfigure}
\par\medskip
\begin{subfigure}[t]{0.49\linewidth}
\centering
\includegraphics[width=\linewidth]{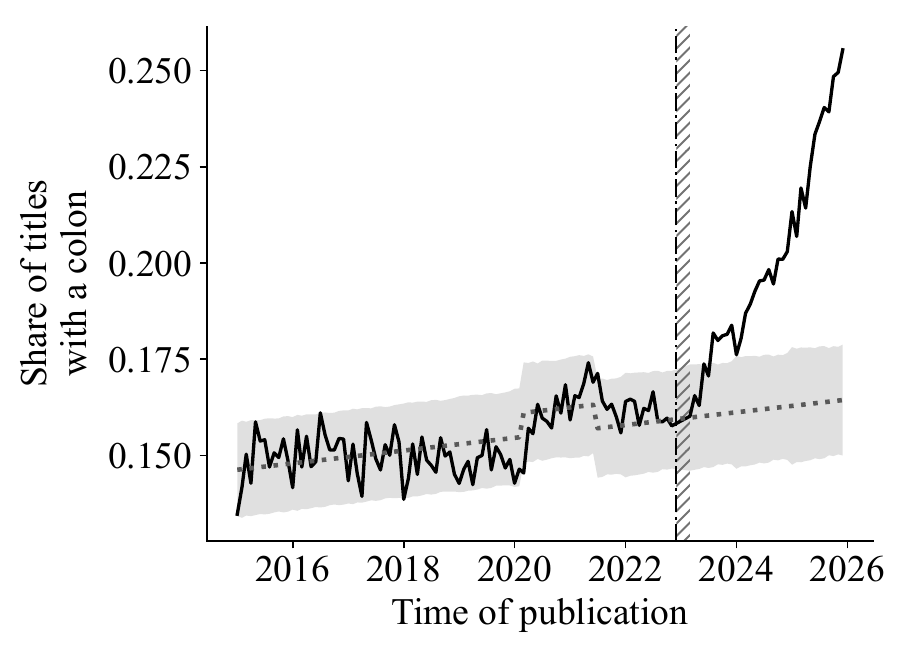}
\caption{Titles with a colon}
\label{fig:meta-colon}
\end{subfigure}\hfill
\begin{subfigure}[t]{0.49\linewidth}
\centering
\includegraphics[width=\linewidth]{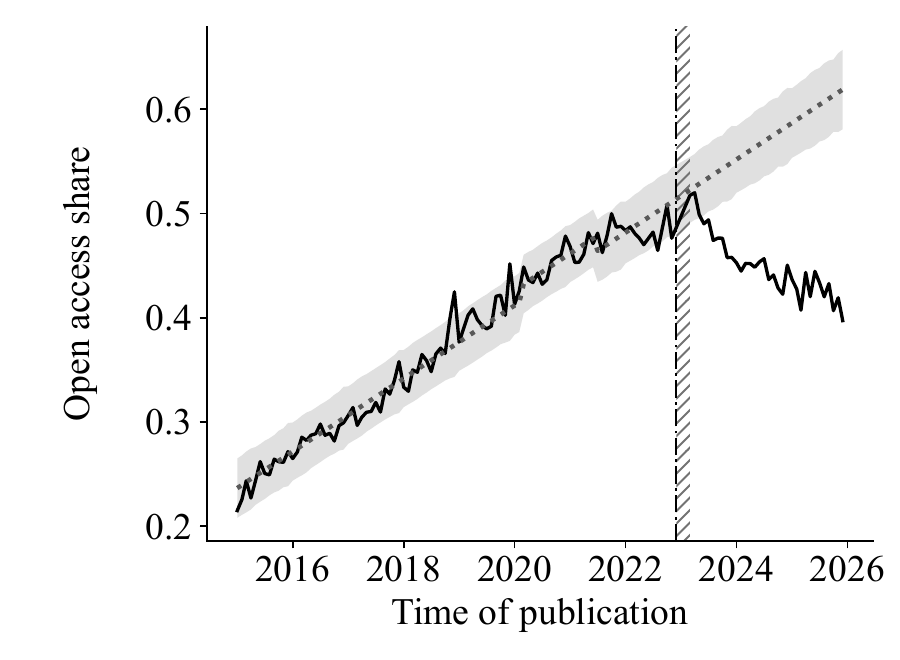}
\caption{Open access share}
\label{fig:meta-oa}
\end{subfigure}
\par\medskip
\includegraphics[width=0.95\linewidth]{fig_panels_legend.pdf}
\caption{Bibliographic Measures and the Projections of Their Pre-2023 Trends. The step in each projection from March 2020 to June 2021 is the
pandemic adjustment in \Cref{eq:its}.}
\label{fig:meta}
\end{figure}

If the literature had changed broadly, the bibliographic measures would move with the
textual ones. Most did not (\Cref{tab:its}, \Cref{fig:meta}). Authors per article
(\Cref{fig:meta-authors}), institutions per article, references
(\Cref{fig:meta-refs}), and pages (\Cref{fig:meta-pages}) all stayed within 4\% of
trend, so the changes anticipated in \Cref{sec:introduction} did not appear: author
counts did not fall, and reference lists did not grow. Monthly publication volume was about 70\% higher in 2023--2025 than in
2015--2022, but it grew along its existing trend (\Cref{fig:meta-articles}); its departure of $-3.7\%$ is not significant ($p = 0.81$). This
measure counts published articles only: a surge in submissions offset by more
rejections would look the same, and submission data are not public. Two measures moved
by much more. The share of titles with a colon rose 23.9\% above trend (\Cref{fig:meta-colon}),
in line with the preference of language models for two-part titles noted in
\Cref{sec:methods},
and open access grew more slowly than projected ($-21.2\%$; \Cref{fig:meta-oa}).
Most of that gap reflects a change in which publishers the articles come from. MDPI,
all of whose articles are open access, fell from 29.0\% of the articles in 2022 to
19.4\% in 2025, while Elsevier, where 25.7\% of articles were open access in 2025,
rose from 40.4\% to 46.9\%. Had each publisher kept its 2022 share of the articles,
the open-access share in 2025 would have been 48.7\% rather than 42.4\%, close to
its 2022 level of 48.3\%. Open-access status is also often recorded some time after
publication, which adds to the gap for the newest articles.

The sharp fall in references per article at the end of 2025 (\Cref{fig:meta-refs})
comes from incomplete records rather than shorter reference lists. Records with no
reference list rose from at most 1.8\% of articles in each month from January to
August 2025 to 6.1\% in December, and records with a list averaged 48.9 references in
December against 55.4 to 57.2 earlier in the year. The newest records are still being
completed, which is why 2026 is excluded. Ending the series in August 2025 moves the
references departure from $-3.4\%$ to $-1.7\%$, although the change in trend is
then borderline ($p = 0.049$, against $p = 0.08$ for the full series).

\begin{table}[htbp]
\caption{Departures from Trend for Abstract and Bibliographic Measures.}
\label{tab:its}
\centering
\singlespacing\small
\begin{tabular}{lrrr}
\toprule
Measure & 2015--2022 mean & Trend-change $p$ & Departure \\
\midrule
\multicolumn{4}{l}{\emph{From abstracts}} \\
Marker words per 10,000 words & 23.6 & $<0.001$ & $+81.8\%$ \\
Control words per 10,000 words & 7.0 & $<0.001$ & $-15.6\%$ \\
Qualifying words per 10,000 words & 23.0 & $<0.001$ & $-13.0\%$ \\
Emphatic words per 10,000 words & 17.1 & $<0.001$ & $+20.1\%$ \\
Share of words with nine or more letters & 0.226 & $<0.001$ & $+10.8\%$ \\
Distinct words per 100 words & 69.4 & $<0.001$ & $+4.3\%$ \\
Flesch-Kincaid grade & 16.8 & $<0.001$ & $+5.9\%$ \\
Abstract length in words & 175.7 & $<0.001$ & $-4.0\%$ \\
\midrule
\multicolumn{4}{l}{\emph{From the bibliographic record}} \\
Articles published per month & 10,402 & $0.81$ & $-3.7\%$ \\
Authors per article & 4.80 & $<0.001$ & $+2.3\%$ \\
Institutions per article & 2.32 & $0.11$ & $-3.7\%$ \\
References per article & 45.7 & $0.08$ & $-3.4\%$ \\
Pages per article & 12.3 & $0.19$ & $-3.7\%$ \\
Title length in words & 13.3 & $0.15$ & $-0.2\%$ \\
Share of titles with a colon & 0.154 & $<0.001$ & $+23.9\%$ \\
Open access share & 0.376 & $<0.001$ & $-21.2\%$ \\
\bottomrule
\end{tabular}
\tabnote{Note: Departure is the mean observed value from March 2023 to December 2025
relative to the projection of the 2015--2022 trend. Trend-change $p$ tests $\beta_3$
in \Cref{eq:its}.}
\end{table}

\subsection{Evidence from article bodies}
\label{sec:results-fulltext}

An abstract is a few hundred words written to a fixed brief, so a change there need
not describe the article. To test whether the change reaches the body of the
article, the study measured the full text of a random sample of articles drawn from
the same corpus. The design called for 10,000 articles. Within each year, every
journal's share of the sample matched its share of that year's articles, and the
years from 2023 onward were sampled at a higher rate, about 0.9\% of their articles
against 0.2\% to 0.4\% for 2015 to 2022, because the after-release side covers three
years against eight. Within each journal-year, articles were drawn at random. Many
drawn articles could not be obtained, because publishers block automated downloads
and most articles are behind a subscription. A journal-year left short was refilled
with further random draws from its articles not yet drawn, and then with open-access
articles drawn at random from the same journal and year, which make up 18.2\% of the
sample. Every article obtained and measured was kept, so the sample holds 12,866
articles, 5,396 published before the release and 7,470 from March 2023 onward, from
250 of the corpus's 256 journals and 29 of its 31 publishers. The other six journals,
which together hold 0.5\% of the corpus, have no article in the sample. Three entered
the corpus only after the sample was drawn, when errors in matching journal names
were corrected; one has fewer than 20 articles in every year, too few to receive an
allocation; and for two, none of the drawn articles could be obtained. Legal open copies were located through
Unpaywall and downloaded by a script that identifies itself, follows each site's
robots.txt rules, and does not request sites that refuse automated access; 1,330
free-to-read articles without a reuse license were measured in memory and not kept,
and 1,091 articles were manually downloaded by the author in a web browser. Most
automated copies come from repositories and are often the accepted manuscript rather
than the published version. Before measurement each document was checked: the
article's title had to appear on its first two pages, and supplementary files,
documents whose text could not be extracted, front matter, and documents under 500
words were set aside. These checks removed 1.0\% of the documents obtained
before the release and 0.9\% of those obtained after. Every estimate in
\Cref{tab:fulltext} weights each article by its journal-year's count in the corpus
divided by its count in the sample. The weights undo both the higher sampling rate
after 2022 and the uneven filling of journal-years, so the comparison describes the
literature rather than the sample; the intervals come from resampling journals.

\begin{table}[htbp]
\caption{Article Bodies Before and After the Release.}
\label{tab:fulltext}
\centering
\singlespacing\small
\begin{tabular}{lrrrr}
\toprule
Measure & Before & After & Change & 95\% interval \\
\midrule
Marker words per 10,000 words & 12.1 & 24.1 & $+98.6\%$ & $+85.7$ to $+112.9$ \\
Control words per 10,000 words & 3.7 & 3.9 & $+4.5\%$ & $-2.2$ to $+11.5$ \\
Qualifying words per 10,000 words & 27.9 & 28.5 & $+2.1\%$ & $-2.0$ to $+6.7$ \\
Emphatic words per 10,000 words & 8.6 & 10.0 & $+16.5\%$ & $+8.4$ to $+25.2$ \\
Share of words with nine or more letters & 0.172 & 0.188 & $+9.3\%$ & $+7.7$ to $+11.0$ \\
Distinct words per 100 words & 66.4 & 67.9 & $+2.1\%$ & $+1.8$ to $+2.5$ \\
Body length in words & 6,546 & 7,387 & $+12.9\%$ & $+10.0$ to $+16.1$ \\
Figures per article & 9.1 & 9.5 & $+4.6\%$ & $+1.2$ to $+8.3$ \\
Tables per article & 2.6 & 2.9 & $+10.7\%$ & $+3.6$ to $+17.7$ \\
Numbered equations per article & 10.8 & 10.1 & $-6.6\%$ & $-14.9$ to $+2.9$ \\
\midrule
States that an AI tool was used in writing & 0.00\% & 1.70\% & $+1.7$ points & $+1.3$ to $+2.2$ \\
Mentions a language model in the body & 0.25\% & 3.74\% & $+3.5$ points & $+2.8$ to $+4.3$ \\
\bottomrule
\end{tabular}
\tabnote{Note: Before: published to November 2022; after: from March 2023. Weighted to
the corpus by journal-year; intervals resample journals (1,000 draws).}
\end{table}

Marker words rose 98.6\% in article bodies, against 81.8\% in abstracts, while the
change in control words cannot be distinguished from zero ($+4.5\%$). Body text, about
forty times longer than an abstract and collected by an entirely separate process,
moved the same way. The result holds within journals: comparing each journal with
itself, among the 191 journals with at least five articles on each side, gives
$+94.9\%$. Leaving out the open-access substitutes gives $+104.3\%$, and keeping only
the 2,750 publisher versions gives $+122.9\%$, so neither the substitutes nor the
repository copies produce the result. Raising the title-match and English-share
thresholds of the quality checks moves the estimate by less than half a point
($+98.6\%$ to $+98.7\%$). Emphatic words and long words rose in bodies as in
abstracts. Qualifying words did not fall in bodies ($+2.1\%$, interval including
zero), so that change appears confined to abstracts. Bodies became longer and carried
slightly more figures and tables, in the direction expected if the tools let authors
present more analysis; the change in equations is not distinguishable from zero.

A disclosure statement is the one measure that needs no inference about style: it
records that an article states, outside its reference list, that an AI tool was used
on its text. Three forms count: the standard declaration ``during the preparation of
this work the authors used [tool] in order to [purpose],'' a declaration heading
followed by a statement of use, and any sentence that states use as done, names a
tool, and names a purpose concerning the text, such as its language, grammar, or
readability. Statements of non-use and quoted journal policy do not count, and every
stored document the rule flags was read to confirm it. No article published from
2015 to 2022 carries such a statement; one in 2023 does, and the share reaches 1.1\%
of articles in 2024 and 3.7\% in 2025. These shares count disclosed use only and
understate even that: disclosure is voluntary at many journals, and a statement worded
outside the three forms is missed. They are therefore a lower bound on use. Disclosure
is also concentrated in one publisher. In 2025, 7.0\% of the sampled Elsevier articles
carry a statement, against 2.2\% at MDPI, 1.4\% at IEEE and at Wiley, and none at the
American Chemical Society or the Royal Society of Chemistry; 79 of the 98 disclosing
articles are Elsevier's, and without Elsevier the weighted share in 2025 is 1.3\%.
Elsevier asks authors to place a separate declaration on the use of generative AI in
the manuscript \citep{elsevier2026aipolicy}, so the share largely records where
authors are asked to declare. The same reason
likely explains why the share is far above the 0.1\% that \citet{he2026policies}
found in full texts published from 2023: their copies came from download links in
OpenAlex, which lead mostly to open-access versions, whereas this sample includes
subscription articles, among them Elsevier's. Separately, the share of articles whose
body mentions a language model at all, mostly as a subject of the research, rose
from 0.2\% to 3.7\%.

\section{Discussion}
\label{sec:discussion}

Four features support reading the result as a change in wording. The departure follows eight years in
which the series stayed close to its trend. It is specific: the control words, from
the same register, did not rise, which argues against a general shift
toward formal prose or an artifact of the method. It does not depend on the method or
on any one journal: false release dates produce small departures and no single
journal moves the estimate by more than a few points. Last, it is confined to wording:
how research is organized and credited, as seen in authors, references, and pages, did
not change.

The design shows a departure from trend, not its cause. Editorial or template
changes across publishers, shifts in the mix of research, and the growth of
non-native English authorship would all leave similar traces; the control words make
some of these less likely but do not rule them out. The release date marks the start
of broad public use of such tools rather than of any single product. The result also
describes a literature, not any paper in it: a shift in average frequency is
compatible with a few authors changing a great deal or many changing a little.

The direction and size agree with \citet{kobak2025} for biomedical abstracts. Two
literatures with different conventions, publishers, and data showing a similar change
is more convincing than either alone. Engineering differs in one respect: without a
uniform public abstract index, any text-based result rests on the machine-readable
part of the literature, which here under-represents the largest publisher.

The findings have three implications for measurement and policy. First, monthly
series built from bibliographic records need care with dates: records that carry only
a year are common at some publishers and, left in, create January peaks that resemble
a seasonal pattern in writing. Second, counting mentions of language models in full
text measures research about the tools, not use of them; a disclosure measure must
require a statement of use, and must set aside statements of non-use, which journals
increasingly print, and quoted journal policy. Even so measured, disclosure is a lower bound, because
many journals do not require it. Third, screening individual manuscripts for the words
studied here would be easy to evade, would penalize writers whose style is naturally
formal, and would fall hardest on non-native English writers, whereas tracking a whole
body of writing is cheap and can inform journal policy without judging any one paper.

\section{Limitations}
\label{sec:limitations}

Publication dates stand in for writing dates, which are unavailable, so the measured
change lags and smooths any change in practice. OpenAlex tends to record the earliest
date it finds for a work \citep{kousha2025llms}, which brings the dates closer to
writing but does not close the gap. The abstract measures rest on the
two-thirds of articles with a usable abstract, which under-represent the largest
publisher; if wording changed differently there, the direction of the resulting bias
is unknown. The full-text measures rest on a sample of about 0.7\% 
of the corpus, drawn at random within journal-years but completed in part with
open-access articles, and most copies obtained automatically are repository versions,
often the accepted manuscript rather than the published text; restricted to
publisher versions, the marker estimate is larger, not smaller. The projection extends each series' own history and cannot
separate this change from others occurring at the same time. The marker list is the
author's choice, built mostly from studies of other fields; the control words and the
lists chosen from the data address that choice but do not remove it. Readability uses
an assumed sentence length, and all measures assume English text.

\section{Conclusion}
\label{sec:conclusion}

Engineering abstracts changed in word choice after the end of 2022, and in a specific
way. Words associated with language model output rose 82\% above their own trend,
while control words of the same register did not rise. Qualifying words fell, emphatic and
longer words rose, and the median abstract went from using none of the marker words
to using several. How articles are assembled stayed close to trend. In the full text of a sample of articles, the same words rose by a similar amount,
and statements disclosing the use of AI writing tools went from none to 3.7\%
of articles in 2025. The
change appears in every subject category. What this study cannot establish is the
cause. Natural next steps are full-text measurement at larger scale, dating by
submission where publishers provide it, and the same analysis in other fields that
lack a uniform abstract index.

\section*{CRediT authorship contribution statement}

\textbf{Taner Cokyasar:} Conceptualization, Methodology, Software, Data curation,
Formal analysis, Investigation, Visualization, Writing -- original draft, Writing --
review \& editing.

\section*{Funding}

This research did not receive any specific grant from funding agencies in the public,
commercial, or not-for-profit sectors.

\section*{Declaration of competing interest}

The author declares that he has no known competing financial interests or personal
relationships that could have appeared to influence the work reported in this paper.

\section*{Data availability}

The replication package contains the code that builds the corpus and runs every
analysis, the journal list with every exclusion and correction, the identifiers of all
1,839,962 articles, the record of the full-text sample with its measurements, and the
computed results. It is available at
\url{https://github.com/tcokyasar44/What_changed_in_engineering_writing_after_AI}, and
the version used here is archived on Zenodo \citep{cokyasar2026package}. Bibliographic data
come from OpenAlex \citep{priem2022openalex}, which places its data in the public
domain. Abstract and article text is not redistributed; the code retrieves it again.
The analysis is deterministic for a fixed data snapshot.

\section*{Declaration of generative AI and AI-assisted technologies in the manuscript
preparation process}

During the preparation of this work the author used Claude (Anthropic), with ChatGPT
(OpenAI) and Gemini (Google) as cross-checks, in order to write and check the data
collection and analysis software, draft parts of the manuscript, and verify the
references. After using these tools, the author reviewed and edited the content and takes full responsibility for the content of the published article. No
language model read, summarized, or judged any article in the study: every measurement
comes from deterministic counting software, and rerunning it on the same data returns
the same numbers.

\bibliographystyle{elsarticle-harv}
\bibliography{refs}

\begin{thebibliography}{43}
\expandafter\ifx\csname natexlab\endcsname\relax\def\natexlab#1{#1}\fi
\providecommand{\url}[1]{\texttt{#1}}
\providecommand{\href}[2]{#2}
\providecommand{\path}[1]{#1}
\providecommand{\DOIprefix}{doi:}
\providecommand{\ArXivprefix}{arXiv:}
\providecommand{\URLprefix}{URL: }
\providecommand{\Pubmedprefix}{pmid:}
\providecommand{\doi}[1]{\href{http://dx.doi.org/#1}{\path{#1}}}
\providecommand{\Pubmed}[1]{\href{pmid:#1}{\path{#1}}}
\providecommand{\bibinfo}[2]{#2}
\ifx\xfnm\relax \def\xfnm[#1]{\unskip,\space#1}\fi
\bibitem[{Alani and Jansen(2026)}]{alani2026stylistic}
\bibinfo{author}{Alani, N.H.S.}, \bibinfo{author}{Jansen, B.J.},
  \bibinfo{year}{2026}.
\newblock \bibinfo{title}{Are scientists starting to sound like {ChatGPT}?
  {S}tylistic diffusion in biomedical abstracts in the {LLM} era}.
\newblock \bibinfo{journal}{Machine Learning and Knowledge Extraction}
  \bibinfo{volume}{8}, \bibinfo{pages}{223}.
\newblock \DOIprefix\doi{10.3390/make8080223}.
\bibitem[{Alsudais(2025)}]{alsudais2025readability}
\bibinfo{author}{Alsudais, A.}, \bibinfo{year}{2025}.
\newblock \bibinfo{title}{Exploring the change in scientific readability
  following the release of {ChatGPT}}.
\newblock \bibinfo{journal}{Journal of Informetrics} \bibinfo{volume}{19},
  \bibinfo{pages}{101679}.
\newblock \DOIprefix\doi{10.1016/j.joi.2025.101679}.
\bibitem[{Bao et~al.(2025)Bao, Zhao, Mao and Zhang}]{bao2025shifts}
\bibinfo{author}{Bao, T.}, \bibinfo{author}{Zhao, Y.}, \bibinfo{author}{Mao,
  J.}, \bibinfo{author}{Zhang, C.}, \bibinfo{year}{2025}.
\newblock \bibinfo{title}{Examining linguistic shifts in academic writing
  before and after the launch of {ChatGPT}: A study on preprint papers}.
\newblock \bibinfo{journal}{Scientometrics} \bibinfo{volume}{130},
  \bibinfo{pages}{3597--3627}.
\newblock \DOIprefix\doi{10.1007/s11192-025-05341-y}.
\bibitem[{Bornmann and Mutz(2015)}]{bornmann2015growth}
\bibinfo{author}{Bornmann, L.}, \bibinfo{author}{Mutz, R.},
  \bibinfo{year}{2015}.
\newblock \bibinfo{title}{Growth rates of modern science: A bibliometric
  analysis based on the number of publications and cited references}.
\newblock \bibinfo{journal}{Journal of the Association for Information Science
  and Technology} \bibinfo{volume}{66}, \bibinfo{pages}{2215--2222}.
\newblock \DOIprefix\doi{10.1002/asi.23329}.
\bibitem[{Cabanac et~al.(2021)Cabanac, Labb\'{e} and
  Magazinov}]{cabanac2021tortured}
\bibinfo{author}{Cabanac, G.}, \bibinfo{author}{Labb\'{e}, C.},
  \bibinfo{author}{Magazinov, A.}, \bibinfo{year}{2021}.
\newblock \bibinfo{title}{Tortured phrases: A dubious writing style emerging in
  science. evidence of critical issues affecting established journals}.
\newblock \bibinfo{howpublished}{arXiv (Cornell University)}.
\newblock \DOIprefix\doi{10.48550/arxiv.2107.06751}. \bibinfo{note}{preprint}.
\bibitem[{Cokyasar(2026)}]{cokyasar2026package}
\bibinfo{author}{Cokyasar, T.}, \bibinfo{year}{2026}.
\newblock \bibinfo{title}{Replication package: What changed in engineering
  writing after generative {AI} chat assistants}.
\newblock \bibinfo{howpublished}{Zenodo}.
\newblock \DOIprefix\doi{10.5281/zenodo.24048694}. \bibinfo{note}{software,
  version 1.0.7}.
\bibitem[{{Committee on Publication Ethics}(2023)}]{cope2023ai}
\bibinfo{author}{{Committee on Publication Ethics}}, \bibinfo{year}{2023}.
\newblock \bibinfo{title}{Authorship and {AI} tools: {COPE} position
  statement}.
\newblock
  \bibinfo{howpublished}{\url{https://publicationethics.org/cope-position-statements/ai-author}}.
\bibitem[{Conroy(2023)}]{conroy2023}
\bibinfo{author}{Conroy, G.}, \bibinfo{year}{2023}.
\newblock \bibinfo{title}{Scientific sleuths spot dishonest {ChatGPT} use in
  papers}.
\newblock \bibinfo{journal}{Nature} \DOIprefix\doi{10.1038/d41586-023-02477-w}.
\bibitem[{Dergaa et~al.(2023)Dergaa, Chamari, Zmijewski and
  Ben~Saad}]{dergaa2023}
\bibinfo{author}{Dergaa, I.}, \bibinfo{author}{Chamari, K.},
  \bibinfo{author}{Zmijewski, P.}, \bibinfo{author}{Ben~Saad, H.},
  \bibinfo{year}{2023}.
\newblock \bibinfo{title}{From human writing to artificial intelligence
  generated text: examining the prospects and potential threats of {ChatGPT} in
  academic writing}.
\newblock \bibinfo{journal}{Biology of Sport} \bibinfo{volume}{40},
  \bibinfo{pages}{615--622}.
\newblock \DOIprefix\doi{10.5114/biolsport.2023.125623}.
\bibitem[{Else(2023)}]{else2023abstracts}
\bibinfo{author}{Else, H.}, \bibinfo{year}{2023}.
\newblock \bibinfo{title}{Abstracts written by {ChatGPT} fool scientists}.
\newblock \bibinfo{journal}{Nature} \bibinfo{volume}{613},
  \bibinfo{pages}{423--423}.
\newblock \DOIprefix\doi{10.1038/d41586-023-00056-7}.
\bibitem[{{Elsevier}(2026)}]{elsevier2026aipolicy}
\bibinfo{author}{{Elsevier}}, \bibinfo{year}{2026}.
\newblock \bibinfo{title}{Generative {AI} policies for journals}.
\newblock
  \bibinfo{howpublished}{\url{https://www.elsevier.com/about/policies-and-standards/generative-ai-policies-for-journals}}.
\newblock \bibinfo{note}{Accessed November 8, 2026}.
\bibitem[{Flesch(1948)}]{flesch1948}
\bibinfo{author}{Flesch, R.}, \bibinfo{year}{1948}.
\newblock \bibinfo{title}{A new readability yardstick}.
\newblock \bibinfo{journal}{Journal of Applied Psychology}
  \bibinfo{volume}{32}, \bibinfo{pages}{221--233}.
\newblock \DOIprefix\doi{10.1037/h0057532}.
\bibitem[{Forchino and Torres-Salinas(2026)}]{forchino2026openalex}
\bibinfo{author}{Forchino, M.V.}, \bibinfo{author}{Torres-Salinas, D.},
  \bibinfo{year}{2026}.
\newblock \bibinfo{title}{The {OpenAlex} database in review: Evaluating its
  applications, capabilities, and limitations}.
\newblock \bibinfo{journal}{Journal of Informetrics} \bibinfo{volume}{20},
  \bibinfo{pages}{101800}.
\newblock \DOIprefix\doi{10.1016/j.joi.2026.101800}.
\bibitem[{Geng and Trotta(2025)}]{geng2025coevolution}
\bibinfo{author}{Geng, M.}, \bibinfo{author}{Trotta, R.}, \bibinfo{year}{2025}.
\newblock \bibinfo{title}{Human-{LLM} coevolution: Evidence from academic
  writing}, in: \bibinfo{booktitle}{Findings of the Association for
  Computational Linguistics: ACL 2025}, \bibinfo{publisher}{Association for
  Computational Linguistics}, \bibinfo{address}{Vienna, Austria}. pp.
  \bibinfo{pages}{12689--12696}.
\newblock \DOIprefix\doi{10.18653/v1/2025.findings-acl.657}.
\bibitem[{{Google Scholar}(2025)}]{scholarmetrics}
\bibinfo{author}{{Google Scholar}}, \bibinfo{year}{2025}.
\newblock \bibinfo{title}{{Google Scholar Metrics}: Engineering and computer
  science}.
\newblock
  \bibinfo{howpublished}{\url{https://scholar.google.com/citations?view_op=top_venues&hl=en}}.
\bibitem[{Gorraiz(2025)}]{gorraiz2025invisible}
\bibinfo{author}{Gorraiz, J.}, \bibinfo{year}{2025}.
\newblock \bibinfo{title}{Acknowledging the new invisible colleague: Addressing
  the recognition of {Open AI} contributions in scientific publishing}.
\newblock \bibinfo{journal}{Journal of Informetrics} \bibinfo{volume}{19},
  \bibinfo{pages}{101642}.
\newblock \DOIprefix\doi{10.1016/j.joi.2025.101642}.
\bibitem[{Haman and \v{S}koln\'{i}k(2024)}]{haman2024}
\bibinfo{author}{Haman, M.}, \bibinfo{author}{\v{S}koln\'{i}k, M.},
  \bibinfo{year}{2024}.
\newblock \bibinfo{title}{Using {ChatGPT} to conduct a literature review}.
\newblock \bibinfo{journal}{Accountability in Research} \bibinfo{volume}{31},
  \bibinfo{pages}{1244--1246}.
\newblock \DOIprefix\doi{10.1080/08989621.2023.2185514}.
\bibitem[{Hartley(2007)}]{hartley2007colons}
\bibinfo{author}{Hartley, J.}, \bibinfo{year}{2007}.
\newblock \bibinfo{title}{Planning that title: Practices and preferences for
  titles with colons in academic articles}.
\newblock \bibinfo{journal}{Library \& Information Science Research}
  \bibinfo{volume}{29}, \bibinfo{pages}{553--568}.
\newblock \DOIprefix\doi{10.1016/j.lisr.2007.05.002}.
\bibitem[{He and Bu(2026)}]{he2026policies}
\bibinfo{author}{He, Y.}, \bibinfo{author}{Bu, Y.}, \bibinfo{year}{2026}.
\newblock \bibinfo{title}{Academic journals' {AI} policies fail to curb the
  surge in {AI}-assisted academic writing}.
\newblock \bibinfo{journal}{Proceedings of the National Academy of Sciences}
  \bibinfo{volume}{123}, \bibinfo{pages}{e2526734123}.
\newblock \DOIprefix\doi{10.1073/pnas.2526734123}.
\bibitem[{Hosseini et~al.(2024)Hosseini, Rasmussen and Resnik}]{hosseini2024}
\bibinfo{author}{Hosseini, M.}, \bibinfo{author}{Rasmussen, L.M.},
  \bibinfo{author}{Resnik, D.B.}, \bibinfo{year}{2024}.
\newblock \bibinfo{title}{Using {AI} to write scholarly publications}.
\newblock \bibinfo{journal}{Accountability in Research} \bibinfo{volume}{31},
  \bibinfo{pages}{715--723}.
\newblock \DOIprefix\doi{10.1080/08989621.2023.2168535}.
\bibitem[{Hubarava and Ineichen(2026)}]{hubarava2026drift}
\bibinfo{author}{Hubarava, H.}, \bibinfo{author}{Ineichen, B.V.},
  \bibinfo{year}{2026}.
\newblock \bibinfo{title}{Accelerated drift of scientific prose in the {LLM}
  era: {E}vidence from an interrupted time-series analysis}.
\newblock \bibinfo{howpublished}{arXiv (Cornell University)}.
\newblock \DOIprefix\doi{10.48550/arXiv.2610.26008}. \bibinfo{note}{preprint}.
\bibitem[{Hyland(1998)}]{hyland1998}
\bibinfo{author}{Hyland, K.}, \bibinfo{year}{1998}.
\newblock \bibinfo{title}{Boosting, hedging and the negotiation of academic
  knowledge}.
\newblock \bibinfo{journal}{Text -- Interdisciplinary Journal for the Study of
  Discourse} \bibinfo{volume}{18}.
\newblock \DOIprefix\doi{10.1515/text.1.1998.18.3.349}.
\bibitem[{Igras-Cybulska and Cybulski(2026)}]{igras2026colons}
\bibinfo{author}{Igras-Cybulska, M.}, \bibinfo{author}{Cybulski, A.},
  \bibinfo{year}{2026}.
\newblock \bibinfo{title}{Did {LLMs} colonize scientific titles? {F}ifty years
  of titular colonicity}.
\newblock \bibinfo{howpublished}{Research Square}.
\newblock \DOIprefix\doi{10.21203/rs.3.rs-11137993/v1}.
  \bibinfo{note}{preprint}.
\bibitem[{Kobak et~al.(2025)Kobak, Gonz\'{a}lez-M\'{a}rquez, Horv\'{a}t and
  Lause}]{kobak2025}
\bibinfo{author}{Kobak, D.}, \bibinfo{author}{Gonz\'{a}lez-M\'{a}rquez, R.},
  \bibinfo{author}{Horv\'{a}t, E.A.}, \bibinfo{author}{Lause, J.},
  \bibinfo{year}{2025}.
\newblock \bibinfo{title}{Delving into {LLM}-assisted writing in biomedical
  publications through excess vocabulary}.
\newblock \bibinfo{journal}{Science Advances} \bibinfo{volume}{11}.
\newblock \DOIprefix\doi{10.1126/sciadv.adt3813}.
\bibitem[{Kousha and Thelwall(2026)}]{kousha2025llms}
\bibinfo{author}{Kousha, K.}, \bibinfo{author}{Thelwall, M.},
  \bibinfo{year}{2026}.
\newblock \bibinfo{title}{How much are {LLMs} changing the language of academic
  papers after {ChatGPT}? {A} multi-database and full text analysis}.
\newblock \bibinfo{journal}{Scientometrics}
  \DOIprefix\doi{10.1007/s11192-026-05601-5}.
\bibitem[{Lewison and Hartley(2005)}]{lewison2005title}
\bibinfo{author}{Lewison, G.}, \bibinfo{author}{Hartley, J.},
  \bibinfo{year}{2005}.
\newblock \bibinfo{title}{What's in a title? {N}umbers of words and the
  presence of colons}.
\newblock \bibinfo{journal}{Scientometrics} \bibinfo{volume}{63},
  \bibinfo{pages}{341--356}.
\newblock \DOIprefix\doi{10.1007/s11192-005-0216-0}.
\bibitem[{Liang et~al.(2024)Liang, Izzo, Zhang, Lepp, Cao, Zhao, Chen, Ye, Liu,
  Huang, McFarland and Zou}]{liang2024reviews}
\bibinfo{author}{Liang, W.}, \bibinfo{author}{Izzo, Z.},
  \bibinfo{author}{Zhang, Y.}, \bibinfo{author}{Lepp, H.},
  \bibinfo{author}{Cao, H.}, \bibinfo{author}{Zhao, X.}, \bibinfo{author}{Chen,
  L.}, \bibinfo{author}{Ye, H.}, \bibinfo{author}{Liu, S.},
  \bibinfo{author}{Huang, Z.X.}, \bibinfo{author}{McFarland, D.A.},
  \bibinfo{author}{Zou, J.}, \bibinfo{year}{2024}.
\newblock \bibinfo{title}{Monitoring {AI}-modified content at scale: A case
  study on the impact of {ChatGPT} on {AI} conference peer reviews}.
\newblock \bibinfo{howpublished}{arXiv (Cornell University)}.
\newblock \DOIprefix\doi{10.48550/arxiv.2403.07183}. \bibinfo{note}{preprint}.
\bibitem[{Liang et~al.(2023)Liang, Yuksekgonul, Mao, Wu and
  Zou}]{liang2023bias}
\bibinfo{author}{Liang, W.}, \bibinfo{author}{Yuksekgonul, M.},
  \bibinfo{author}{Mao, Y.}, \bibinfo{author}{Wu, E.}, \bibinfo{author}{Zou,
  J.}, \bibinfo{year}{2023}.
\newblock \bibinfo{title}{{GPT} detectors are biased against non-native
  {English} writers}.
\newblock \bibinfo{journal}{Patterns} \bibinfo{volume}{4},
  \bibinfo{pages}{100779}.
\newblock \DOIprefix\doi{10.1016/j.patter.2023.100779}.
\bibitem[{Liang et~al.(2025)Liang, Zhang, Wu, Lepp, Ji, Zhao, Cao, Liu, He,
  Huang, Yang, Potts, Manning and Zou}]{liang2024papers}
\bibinfo{author}{Liang, W.}, \bibinfo{author}{Zhang, Y.}, \bibinfo{author}{Wu,
  Z.}, \bibinfo{author}{Lepp, H.}, \bibinfo{author}{Ji, W.},
  \bibinfo{author}{Zhao, X.}, \bibinfo{author}{Cao, H.}, \bibinfo{author}{Liu,
  S.}, \bibinfo{author}{He, S.}, \bibinfo{author}{Huang, Z.},
  \bibinfo{author}{Yang, D.}, \bibinfo{author}{Potts, C.},
  \bibinfo{author}{Manning, C.D.}, \bibinfo{author}{Zou, J.},
  \bibinfo{year}{2025}.
\newblock \bibinfo{title}{Quantifying large language model usage in scientific
  papers}.
\newblock \bibinfo{journal}{Nature Human Behaviour} \bibinfo{volume}{9},
  \bibinfo{pages}{2599--2609}.
\newblock \DOIprefix\doi{10.1038/s41562-025-02273-8}.
\bibitem[{Lopez~Bernal et~al.(2016)Lopez~Bernal, Cummins and
  Gasparrini}]{bernal2017its}
\bibinfo{author}{Lopez~Bernal, J.}, \bibinfo{author}{Cummins, S.},
  \bibinfo{author}{Gasparrini, A.}, \bibinfo{year}{2016}.
\newblock \bibinfo{title}{Interrupted time series regression for the evaluation
  of public health interventions: a tutorial}.
\newblock \bibinfo{journal}{International Journal of Epidemiology}
  \DOIprefix\doi{10.1093/ije/dyw098}.
\bibitem[{Newey and West(1987)}]{newey1987}
\bibinfo{author}{Newey, W.K.}, \bibinfo{author}{West, K.D.},
  \bibinfo{year}{1987}.
\newblock \bibinfo{title}{A simple, positive semi-definite, heteroskedasticity
  and autocorrelation consistent covariance matrix}.
\newblock \bibinfo{journal}{Econometrica} \bibinfo{volume}{55},
  \bibinfo{pages}{703}.
\newblock \DOIprefix\doi{10.2307/1913610}.
\bibitem[{Ni and Wang(2025)}]{ni2025disruptions}
\bibinfo{author}{Ni, R.}, \bibinfo{author}{Wang, J.}, \bibinfo{year}{2025}.
\newblock \bibinfo{title}{Navigating disruptions: The effects of the pandemic
  on scientific collaboration and research novelty in {Hong Kong}}.
\newblock \bibinfo{journal}{Journal of Informetrics} \bibinfo{volume}{19},
  \bibinfo{pages}{101656}.
\newblock \DOIprefix\doi{10.1016/j.joi.2025.101656}.
\bibitem[{{OpenAI}(2022)}]{openai2022chatgpt}
\bibinfo{author}{{OpenAI}}, \bibinfo{year}{2022}.
\newblock \bibinfo{title}{Introducing {ChatGPT}}.
\newblock \bibinfo{howpublished}{\url{https://openai.com/blog/chatgpt}}.
\bibitem[{Priem et~al.(2022)Priem, Piwowar, Barrett, Demes, McCafferty, Meyer,
  Orr and Portenoy}]{priem2022openalex}
\bibinfo{author}{Priem, J.R.}, \bibinfo{author}{Piwowar, H.},
  \bibinfo{author}{Barrett, J.}, \bibinfo{author}{Demes, K.},
  \bibinfo{author}{McCafferty, N.}, \bibinfo{author}{Meyer, C.},
  \bibinfo{author}{Orr, R.}, \bibinfo{author}{Portenoy, J.},
  \bibinfo{year}{2022}.
\newblock \bibinfo{title}{{OpenAlex} snapshot}.
\newblock \bibinfo{howpublished}{arXiv (Cornell University)}.
\newblock \DOIprefix\doi{10.48550/arxiv.2205.01833}. \bibinfo{note}{preprint}.
\bibitem[{Sebo et~al.(2026)Sebo, Nie and Wang}]{sebo2026titles}
\bibinfo{author}{Sebo, P.}, \bibinfo{author}{Nie, B.}, \bibinfo{author}{Wang,
  T.}, \bibinfo{year}{2026}.
\newblock \bibinfo{title}{How {ChatGPT} writes scientific titles in medical
  research: Structural and content differences compared to human authors}.
\newblock \bibinfo{journal}{Journal of the Medical Library Association}
  \bibinfo{volume}{114}, \bibinfo{pages}{222--237}.
\newblock \DOIprefix\doi{10.5195/jmla.2026.2266}.
\bibitem[{{Springer Nature}(2026)}]{springernature2026state}
\bibinfo{author}{{Springer Nature}}, \bibinfo{year}{2026}.
\newblock \bibinfo{title}{The state of {AI} use among researchers}.
\newblock
  \bibinfo{howpublished}{\url{https://group.springernature.com/br/group/media/press-releases/early-career-researchers-lead-shift-towards-ai-first-discovery/52988132}}.
\newblock \bibinfo{note}{White paper with Fudan University and the Shanghai
  Academy of AI for Science; survey of 10,480 researchers fielded in March
  2026}.
\bibitem[{{Stanford Institute for Human-Centered Artificial
  Intelligence}(2026)}]{aiindex2026}
\bibinfo{author}{{Stanford Institute for Human-Centered Artificial
  Intelligence}}, \bibinfo{year}{2026}.
\newblock \bibinfo{title}{The 2026 {AI} index report}.
\newblock
  \bibinfo{howpublished}{\url{https://hai.stanford.edu/ai-index/2026-ai-index-report}}.
\bibitem[{Stokel-Walker(2023)}]{stokelwalker2023}
\bibinfo{author}{Stokel-Walker, C.}, \bibinfo{year}{2023}.
\newblock \bibinfo{title}{{ChatGPT} listed as author on research papers: many
  scientists disapprove}.
\newblock \bibinfo{journal}{Nature} \bibinfo{volume}{613},
  \bibinfo{pages}{620--621}.
\newblock \DOIprefix\doi{10.1038/d41586-023-00107-z}.
\bibitem[{Sun et~al.(2023)Sun, Liu, Li and Ma}]{sun2023expedited}
\bibinfo{author}{Sun, Z.}, \bibinfo{author}{Liu, S.}, \bibinfo{author}{Li, Y.},
  \bibinfo{author}{Ma, C.}, \bibinfo{year}{2023}.
\newblock \bibinfo{title}{Expedited editorial decision in {COVID-19} pandemic}.
\newblock \bibinfo{journal}{Journal of Informetrics} \bibinfo{volume}{17},
  \bibinfo{pages}{101382}.
\newblock \DOIprefix\doi{10.1016/j.joi.2023.101382}.
\bibitem[{Thorp(2023)}]{thorp2023}
\bibinfo{author}{Thorp, H.H.}, \bibinfo{year}{2023}.
\newblock \bibinfo{title}{{ChatGPT} is fun, but not an author}.
\newblock \bibinfo{journal}{Science} \bibinfo{volume}{379},
  \bibinfo{pages}{313--313}.
\newblock \DOIprefix\doi{10.1126/science.adg7879}.
\bibitem[{Van~Noorden and Perkel(2023)}]{vannoorden2023}
\bibinfo{author}{Van~Noorden, R.}, \bibinfo{author}{Perkel, J.M.},
  \bibinfo{year}{2023}.
\newblock \bibinfo{title}{{AI} and science: what 1,600 researchers think}.
\newblock \bibinfo{journal}{Nature} \bibinfo{volume}{621},
  \bibinfo{pages}{672--675}.
\newblock \DOIprefix\doi{10.1038/d41586-023-02980-0}.
\bibitem[{Wagner et~al.(2002)Wagner, Soumerai, Zhang and
  Ross-Degnan}]{wagner2002segmented}
\bibinfo{author}{Wagner, A.K.}, \bibinfo{author}{Soumerai, S.B.},
  \bibinfo{author}{Zhang, F.}, \bibinfo{author}{Ross-Degnan, D.},
  \bibinfo{year}{2002}.
\newblock \bibinfo{title}{Segmented regression analysis of interrupted time
  series studies in medication use research}.
\newblock \bibinfo{journal}{Journal of Clinical Pharmacy and Therapeutics}
  \bibinfo{volume}{27}, \bibinfo{pages}{299--309}.
\newblock \DOIprefix\doi{10.1046/j.1365-2710.2002.00430.x}.
\bibitem[{{Wiley}(2025)}]{wiley2025explanaitions}
\bibinfo{author}{{Wiley}}, \bibinfo{year}{2025}.
\newblock \bibinfo{title}{{AI} adoption jumps to 84\% among researchers as
  expectations undergo significant ``reality check''}.
\newblock
  \bibinfo{howpublished}{\url{https://newsroom.wiley.com/press-releases/press-release-details/2025/AI-Adoption-Jumps-to-84-Among-Researchers-as-Expectations-Undergo-Significant-Reality-Check/default.aspx}}.
\newblock \bibinfo{note}{Press release on the ExplanAItions survey of 2,430
  researchers}.

\end{thebibliography}

\end{document}